\def\tr{\hbox{tr}}
\documentstyle[12pt,aps]{revtex}

\begin{document}
\draft
\title{On the standard model and parity conservation}
\author{She-Sheng Xue
}
\address{
ICRA, INFN  and
Physics Department, University of Rome ``La Sapienza", 00185 Rome, Italy
}


\maketitle

\centerline{xue@icra.it}

\begin{abstract}

Motivated by the theoretical paradox between the parity-violating gauge symmetries 
of the standard model and the fundamental regulator, we propose that
the extension of the standard model to the high-energy region should be made by adding
effective high-dimension operators whose dynamics gives an infrared 
scaling region, where not only the standard model appears as an asymptotic 
chiral-gauge theory, but also vectorlike one-particle-irreducible 
functions are induced to describe ``new physics''. We formulate the standard model on a
lattice by introducing a right-handed neutrino and a sterile left-handed neutrino 
to form the high-dimension operators. Analyzing spectra and one-particle-irreducible 
functions induced by these high-dimension operators, we find that in the low-energy region,
the parity-violating standard model is consistently defined, 
while the parity-conservation gauge-symmetry is restored in the high-energy region. 

\end{abstract}

\pacs{
11.15Ha,
11.30.Rd, 
11.30.Qc
}

\newpage

\narrowtext

\section{\it Introduction.}\label{introduction}

The parity-violating feature of elementary particle physics in the 
low-energy region is strongly phenomenologically supported. On the 
basis of this feature and the gauge principle, the successful standard model (SM) 
for elementary particle physics is constructed in the form of a renormalizable 
quantum field theory with chiral (parity-violating) gauge symmetries. The SM is 
very successful in particularly describing gauge fields, fermion fields, and 
their couplings in the 
low-energy region. As a renormalizable quantum field theory, the SM has to be properly 
regularized and quantized in such a way that the relevant spectra and 
one-particle-irreducible (1PI) functions
attributed to low-energy modes are precisely defined, while irrelevant contributions 
from high-energy modes are renormalized away. Though the triumph of the SM is 
demonstrated by low-energy experiments, theoretical inconsistency and paradox 
still confuse theoretical particle physicists today.
Apart from the embarrassing inconsistency of the quadratic divergence in the Higgs sector, 
for which an elegant supersymmetry is introduced, the theoretical paradox of regularizing 
a chiral gauge theory in short-distances - high-energy-region- 
is one of most important issues in the theoretical particle physics.

With very generic axioms of quantum field theories, Nielsen and Ninomiya mathematically 
demonstrate the ``no-go" theorem 
\cite{nn81} that the quantum field theories with chiral gauge symmetries, 
like the SM, cannot be consistently 
regularized on the lattice for either gauge symmetry-breakings (giving up the gauge
principle) or vectorlike fermion-doubling phenomenon (contradicting low-energy phenomenon). 
This paradox is in fact generic and independent of regularization 
schemes\cite{nn81plb,nn91}. As a consequence, this paradox, either 
giving up the gauge principle or contradicting the low-energy phenomenon, seems unavoidable. 

On the other hand, in the high-energy region, the very-small-scale structure of the 
space-time at distances of the 
Planck length, where violent fluctuations of the quantum gravity take 
place, can exhibit rather complex structure of a space-time ``string" 
or ``foam", instead of a simple space-time point. As the consequence 
of these fundamental constituents of the space-time, the physical 
space-time, the arena of physical reality, gets endowed with a fundamental 
length and the gauge-invariant fundamental theory must be {\it finite}. This implies that the 
quantum gravity could play a role of the nature regulator at the Planck length 
for the fundamental theory with chiral gauge interactions between gauge 
and fermion fields, e.g., the $SU(5)$-theory. In the view of the fundamental 
theory of particle physics being finite, e.g., the string theory, the paradox raised by 
the ``no-go" theorem is even more profound and far-reaching.  
 
One might regard that such an inconsistency should be a technicality of 
regularization schemes and would not intrinsically jeopardize the fundamental theory for
the following reasons: (i) the vectorlike fermion-doubling is
an artificial defect of the lattice-regularization; (ii) the non-local gauge 
anomalies, resulted from chiral gauge-symmetry-breakings by regularizations, do 
not receive renormalization from high-order perturbative contributions and are 
completely canceled within the fermion content of the SM; (iii) local chiral 
gauge-symmetry-breaking terms can be {\it self-consistently} eliminated by 
adding local counterterms and fine-tuning order by order in perturbation theories in terms of
small and smooth gauge fields. The third point is actually originated from 
our experience of dealing with a perturbative vectorlike gauge theory (like QED) by a 
gauge variant regularization scheme.

The third point for QED is indeed {\it self-consistent} because the vectorlike 
gauge-symmetry and 
relative Ward identities are intrinsically preserved by the regularization and local
gauge-symmetry-breakings due to an inadequate choice of gauge variant regularizations 
are indeed artificial. While, such a {\it self-consistency} is not so obvious and 
justified for perturbative chiral-gauge theories, since we are not guaranteed 
by an exactly gauge-invariant regularization due to the ``no-go'' theorem. 
On the other hand, for the non-perturbative and non-smooth
variations of gauge fields at short distances, local gauge-symmetry-breakings cannot be 
eliminated by a fine-tuning procedure without exact gauge symmetries. Beside, it is rather 
unnatural that the gauge principle is preserved by a fine-tuning procedure. 
We have to first get around the ``no-go'' theorem to find a gauge invariant regularization 
scheme, not only for approaching an asymptotical chiral-gauge theory of perturbative and smooth 
gauge fields at long distances, but also for non-perturbatively regularizing chiral 
gauge theories at short distances. Once this can be achieved or proved, 
the exact gauge-symmetry and Ward identities guarantee the self-consistency of 
eliminating local gauge-symmetry-breakings by adding and fine-tuning appropriate 
counterterms in perturbative calculations, as the normal renormalization prescription 
of a gauge-invariant quantum field theory.    

A great effort\cite{gw}-\cite{11112} had been made for finding a gauge-invariant, both 
perturbative and non-perturbative regularization (lattice) of chiral gauge theories, 
since the ``no-go'' theorem was proposed in 1981. For a long time, the general belief 
has been that the gauge-invariant regularization of chiral 
gauge theories seems to be impossible. 
We have been in fact in the position of opposing to this opinion for a decade (see 
Xue's reference). The reasons for us 
{\it a priori} to take this position in the early time are following:  

\begin{itemize}

\item{(i)}
the general belief has never been demonstrated. 
While, in practice, the Rome approach\cite{rome,terev} shows a complete {\it self-consistent} 
framework where gauge-symmetry-breakings due to the Wilson fermion\cite{wilson} are canceled by 
adding gauge variant counterterms to force the Ward identities of chiral gauge 
symmetries to be held. This inspires us, in 
principle, there must be a universal class of gauge invariant regularizations for chiral 
gauge theories, otherwise the Rome approach would not be self-consistent. 
Such gauge invariant regularizations are most probably realized by a dynamics, 
instead of fine-tuning symmetries; 
   
\item{(ii)}
since the gauge principle must be true, we might be allowed to take an 
attitude that the paradox could imply a hint of ``new physics'' beyond 
the SM\cite{xue90}. In fact, one of prerequisites of the ``no-go'' theorem is that the action 
of chiral gauge theories is bilinear in fermion fields. This indicates that ``new physics'' 
would be effectively represented by gauge-invariant high-dimension 
operators of fermion fields in the high-energy region. 
Searching for such operators to non-perturbatively regularize the SM on the 
lattice is not only for achieving the SM as 
an asymptotic chiral gauge theory in the low-energy region, but also for finding 
relevant spectra and 1PI functions in the high-energy region, which 
sheds light on what would possibly be ``new physics''.
 
\end{itemize}
These effective high-dimension operators at the lattice scale $1/a$ are possibly 
induced from the fundamental theory with the quantum gravity included at the Planck scale, 
for instance, the string theory. The use of the lattice regularization is not essential 
and we can use other gauge-invariant regularizations at the scale $1/a$, since the physical 
relevant 1PI functions in the infrared scaling region are determined up to some irrelevant 
and vanishing high-dimension terms.

Searching for a chiral-gauge symmetric approach to properly regularize 
the standard model on the lattice has been greatly challenging to particle 
physicists for the last two decades. The recent development \cite{nerev,lurev} based
on the Ginsparg-Wilson equation\cite{gw}, show an gauge-invariant
manner to regularize perturbation theory, which will be discussed later.
In this paper, we focus on the approach that is the modeling 
by appropriately introducing local and high-dimension interactions of fermion 
fields (or bosons) at the lattice scale\cite{ep}-\cite{px91},
\cite{monrev}-\cite{px} and \cite{xue96}-\cite{mc}. Several universal 
classes of such models on the lattice have been investigated and 
shown to fail\cite{ss,aoki,ahx} and \cite{perev,pgr,shrev}. It is then a 
general belief that the 
phenomenon of spontaneous symmetry-breakings in the intermediate 
coupling region and the argument of anomaly-cancelation within vectorlike 
spectra in the strong-coupling region prevent such modelings from having 
a gauge-invariant low-energy scaling region for chiral gauged fermions. 
Nevertheless, in refs.\cite{xue96,xue97}, an $SU_L(2)$ chiral gauge theory with the 
peculiar local four-fermion interactions was proposed and a gauge-invariant 
low-energy scaling region was advocated. The dynamics of realizing an 
asymptotic $SU_L(2)$ chiral gauge theory in such a low-energy region 
was intensively studied\cite{xue97,xue99,xue00}.

In order to make the model so constructed and its dynamics be more physically 
intuitive and be easily checked, in ref.\cite{11112} we study the (11112) model on the 1+1 
dimensional
lattice with four-fermion interactions analogous to that of the $SU_L(2)$ model 
in refs.\cite{xue97,xue99,xue00}. It is shown that the exact computations of 
relevant 
$S$-matrices demonstrate a loophole that the model and its dynamics 
can possibly evade the ``no-go'' theorem of Nielsen and Ninomiya. The 
low-energy chiral gauged spectra of the (11112) model can be achieved, 
consistently with both the cancelation of gauge anomalies preserving the gauge 
symmetry and flavour-singlet anomalies obeying the index theorem.

In this article, we study the analogous four-fermion interactions
introduced for the fermion content of the SM with two additional 
neutrinos\footnote{A brief report of this study is given in ref.\cite{xuesm}}: 
a right-handed Weyl neutrino $\nu_R\equiv\chi_R$ and a sterile left-handed
Weyl neutrino $\chi_L$. The spectra and 1PI functions 
induced by the four-fermion interactions are analyzed by using Ward identities, 
the strong- and weak-coupling 
analysis from section (\ref{standard}) to section (\ref{scenario}).
On the basis of these spectrum and 1PI functions, we study and 
discuss gauge anomalies and fermion-number (B+L) anomalies in section 
(\ref{sanomaly}). In section (\ref{soft}), we discuss the possible 
relationship between the presented model and the recent approaches\cite{nerev,lurev} based on
the Ginsparg-Wilson equation\cite{gw}, and the possibility of a soft 
spontaneous symmetry-breaking for the fermion mass generation.  
As a result, in sections (\ref{scenario},\ref{restore}), we present 
a whole scenario of the model in the gauge-invariant low-energy scaling region: (i) 
the SM appears as an asymptotic chiral gauge theory with two additional neutrinos; 
(ii) massive and vectorlike spectra and 1PI functions indicate that
the parity-conservation gauge symmetries are restored in the high-energy region.

\section{The standard model}\label{standard}

\vskip0.5cm
\noindent{\it Introduction.}\hskip0.2cm 
We discuss the formulation of the standard model (SM) on the lattice. 
In refs.\cite{xue97,xue99,xue00}, 
we intensively study the same type of the four-fermion interactions for 
an $SU_L(2)$ model in the four 
dimensional space-time, where we consider only one left-handed fermion doublet coupling 
to the $SU_L(2)$ gauge field and one neutral spectator $\chi_R$. In these studies, 
we in particular pay much attention on trying to avoid hard spontaneous symmetry 
breakings in the intermediate values of the four-fermion couplings. The same type of 
four-fermion interactions will be respectively introduced for the both
left-handed sector and right-handed sector of the SM, containing gauged 
doublets and singlets 
of the lepton and quarks in three generations. The dynamics 
of four-fermion interactions for realizing a low-energy spectrum of chiral 
gauged fermions should not be changed by the difference in the fermion 
contents between the SM and previously studied models.

\vskip0.5cm
\noindent{\it Naive action.}\hskip0.3cm 
For being economic in notations and simplifying illustrations, we only explicitly write 
$\psi_L$ for all left-handed fermion doublets of the SM:
\begin{equation}
\psi_L=\left(\matrix{\nu_L\cr e_L}\right),\hskip0.3cm \left(\matrix{u_L\cr d_L}\right);
\cdot\cdot\cdot
\label{sml}
\end{equation}
and $\psi_R$ for all right-handed fermion singlets of the SM:
\begin{equation}
\psi_R=e_R,\hskip0.2cm u_R, \hskip0.2cm d_R;\cdot\cdot\cdot .
\label{smr}
\end{equation}
Both the left-handed doublets $\psi_L$ and the right-handed singlets $\psi_R$ are the 
eigenstates of the gauge-group $SU_c(3)\otimes SU_L(2)\otimes U_Y(1)$ of the SM. 
The Higgs sector is completely disregarded for the time being and all fermion 
fields are massless and two-component Weyl fields. 

The naively regularized action for the SM on 
the four dimensional lattice is given by
\begin{equation}
S_\circ ={1\over 2a}\sum_x\Big(\bar\psi_L(x) D^L_\mu\cdot\gamma_\mu\psi_L(x)
+\bar\psi_R(x) D^R_\mu\cdot\gamma_\mu\psi_R(x)\Big),
\label{sfree}
\end{equation}
which are gauge-invariant kinetic terms for the doublets $\psi_L$ and singlets 
$\psi_R$, whose dimensions are [$a^{1/2}$]. In eq.(\ref{sfree}), $x$ is the 
integer label of four-dimensional space-time sites and the differential 
operators $D_\mu^{L,R}$ are,
\begin{equation} 
D_\mu^{L,R}=([U_\mu(x)]^{L,R}\delta_{x,x+1}
-[U_\mu^\dagger(x)]^{L,R}\delta_{x,x-1}),
\label{skinetic} 
\end{equation} 
where the gauge link variable $U_\mu(x)$ is an element of the 
$SU_c(3)\otimes SU_L(2)\otimes U_Y(1)$ gauge group of the SM. 

This naively regularized action at the lattice scale preserves the parity-violating 
(chiral) gauge symmetries of the SM, however its spectrum at the low-energy scale is 
vectorlike, comprising not only the normal fermions (\ref{sml}) and (\ref{smr}) 
of the SM, but also extra fermion species, as the consequence of the ``no-go'' 
theorem. The extra fermion species(doublers) carrying the 
same quantum charges of the gauge group but opposite chiralities appear as 
low-energy excitations at the edges of the Brillouin zone in the momentum space 
$p=a\tilde p+\pi_A$, where $\pi_A\not=0$ running over fifteen lattice momenta. 
This spectrum does not respect the parity-violating gauge symmetries, inconsistently 
with phenomenological observations. As discussed in the 
introductory section (\ref{introduction}) at the beginning of the paper, on the way of 
finding a resolution to evade the ``no-go'' theorem, we 
could be led to the right track to reach the new physics beyond the SM.  

\vskip0.5cm
\noindent{\it Sterile neutrinos and four-fermion interactions.}\hskip0.3cm 
Analogous to the two neutral ``spectators'' $\chi_L$ and $\chi_R$ 
introduced in the (11112) model\cite{11112}, in the light of the Ockham razor, we 
introduce only two Weyl neutrinos $\chi_L$ and $\chi_R$ for 
three fermion generations of the SM. $\chi_R$ is the right-handed neutrino 
$\chi_R\equiv\nu_R$ and $\chi_L$ is a left-handed sterile neutrino. These two 
Weyl fermions neither carry any quantum charges of the gauge group of the SM, 
nor any quantum numbers of symmetries associating to lepton, baryon 
and flavour(generation) numbers of the SM. The kinetic terms for these two  
sterile neutrinos are given by
\begin{equation}
S_{\rm sterile} ={1\over 2a}\sum_x\Big(\bar\chi_L(x) \partial^\mu
\cdot\gamma_\mu\chi_L(x)+\bar\chi_R(x) \partial^\mu\cdot\gamma_\mu\chi_R(x)\Big),
\label{sterile}
\end{equation}
where $\chi_R$ and $\chi_L$ have the dimension [$a^{1/2}$].
 
At the lattice scale $1/a$, the right-handed neutrino 
$\chi_R$ couples to the left-handed doublets $\psi_L$: 
\begin{equation}
S^L_i=g_1\bar\psi_L(x)\cdot\chi_R(x)\bar\chi_R(x)\cdot\psi_L(x)
+g_2\bar\psi_L(x)\cdot\left[\Delta\chi_R(x)\right]
\left[\Delta\bar\chi_R(x)\right]\cdot\psi_L(x),
\label{shil}
\end{equation}
and the left-handed neutrino $\chi_L$ couples to the 
right-handed singlets $\psi_R$:
\begin{equation}
S^R_i=g_1\bar\psi_R(x)\cdot\chi_L(x)
\bar\chi_L(x)\cdot\psi_R(x)+g_2\bar\psi_R(x)\cdot\left[\Delta\chi_L(x)\right]
\left[\Delta\bar\chi_L(x)\right]\cdot\psi_R(x).
\label{shir}
\end{equation}
The couplings $g_1,g_2$ have the dimension $[a^{-2}]$ and are the same for  
both the left-handed sector $S^L_i$ and the right-handed sector $S^R_i$. 
The differential operator $\Delta$ in the four dimensions is defined as,
\begin{equation}
\Delta\chi_{L,R}(x)\equiv\sum_\mu
\Big[ \chi_{L,R}(x+a_\mu)+\chi_{L,R}(x-a_\mu)-2\chi_{L,R}(x)\Big],
\label{swisf0}
\end{equation}
and its Fourier transformation
\begin{equation}
w(p)={1\over2}\sum_xe^{-ipx}\Delta(x)={1\over2}\sum_\mu
\left(\cos(p_\mu)-1\right).
\label{swisf}
\end{equation}

The total action of the lattice-regularization of the standard 
model (RSM for short) at the lattice scale is then given by
\begin{equation}
S =S_\circ +S_{\rm sterile} + S^L_i + S^R_i,
\label{stotal}
\end{equation}
which exactly preserves the gauge symmetries 
$SU_c(3)\otimes SU_L(2)\otimes U_Y(1)$ of the SM. Since all fermions 
are massless Weyl fermions, the left-handed doublets $\psi_L$ and  
the right-handed singlets $\psi_R$ are completely 
separated from each other in the kinetic actions (\ref{skinetic}) and 
(\ref{sterile}), which indicates the RSM total action (\ref{stotal}) 
can be divided into
the left-handed sector (\ref{sml},\ref{shil}) and right-handed sector 
(\ref{smr},\ref{shir}). The exact gauge symmetries for the 
left-handed sector are 
\begin{equation}
SU^L_c(3)\otimes SU_L(2)\otimes U^L_Y(1)
\label{syl}
\end{equation}
and for the right-handed sector are 
\begin{equation}
SU^R_c(3)\otimes U^R_Y(1),
\label{syr}
\end{equation}
where $U^{L,R}_Y(1)$ 
are the $U_Y(1)$ hypercharges for the left- and right-handed sectors respectively, 
and $SU^{L,R}_c(3)$ are the left- and right-handed 
representations of the $SU_c(3)$ only for left- and right-handed quarks. 

In addition to the 
gauge symmetries, there are global Abelian symmetries in the RSM: (i) the $U_L(1)$ 
symmetry for the number of 
left-handed fermions $\psi_L$ and the $U_R(1)$ symmetry for the number 
of right-handed fermions $\psi_R$, which are related to the lepton and 
baryon numbers of the SM; (ii) the $U_{\chi_R}(1)$ and $U_{\chi_L}(1)$ 
symmetries for two sterile neutrinos
$\chi_R$ and $\chi_L$. Beside, this lattice-regularized action (\ref{stotal}) 
possesses the exact $\chi_R$-shift-symmetry and $\chi_L$-shift-symmetry for the 
four-fermion coupling $g_1\rightarrow 0$, namely, the total action $S$ (\ref{stotal}) 
of the RSM is invariant under the following transformations:
\begin{eqnarray}
\chi_R(x)&\rightarrow& \chi_R(x) +\epsilon_R;\hskip0.3cm 
\bar\chi_R(x)\rightarrow \bar\chi_R(x) +\bar\epsilon_R,
\label{sshiftr}\\
\chi_L(x)&\rightarrow& \chi_L(x) +\epsilon_L;\hskip0.3cm 
\bar\chi_L(x)\rightarrow \bar\chi_L(x) +\bar\epsilon_L,
\label{sshiftl}
\end{eqnarray}
where $\epsilon_L$ and $\epsilon_R$ are space-time 
independent Grassmann variables.

\vskip0.5cm
\noindent{\it The properties of four-fermion interactions.}\hskip0.3cm
These four-fermion interactions $S^L_i$ (\ref{shil}) and $S^R_i$ (\ref{shir}) are (i) 
dimension-6 relevant operators ($g_1$) for both normal fermions of the SM and doublers  
and (ii) dimension-10 relevant operators ($g_2$) only for doublers. 
Concerning only on preserving the gauge symmetries of the SM, we have 
two possibilities of four-fermion interactions $S^L_i$ (\ref{shil}) 
and $S^R_i$ (\ref{shir}) in the RSM at the lattice scale:

\begin{itemize}

\item{(i)} 
one of fermion fields $\bar\psi_L$ and $\psi_L$ ($\bar\psi_R$ and $\psi_R$) 
is from one generation and another is from 
another generation of the SM. The four-fermion coupling is a matrix 
in the flavour space. This explicitly has the flavour-changing effect. 
We do not consider this case in the present paper;

\item{(ii)}
both fermion fields $\bar\psi_L$ and $\psi_L$ ($\bar\psi_R$ and $\psi_R$) 
are from either the lepton-sector or the quark-sector in the same 
generation of the SM. This preserves global symmetries $U_{L,R}(1)$ for 
the baryon- and lepton-numbers and flavour-symmetries of the SM.

\end{itemize}

We can also consider the interaction vertex of t'Hooft type\cite{mc}, 
which is a dimension-6 operator violating the lepton and 
baryon numbers ($B+L$). There are {\it a priori} no 
rules to preclude more complex high-dimension and low-dimension operators 
that process gauge symmetries of the SM and the shift-symmetries. The 
important task is to find a universal class of high-dimension operators at the 
lattice scale, whose dynamics gives rise to a low-energy scaling region for relevant 
spectra and renormalizable operators of the SM. 

\section{Ward identities}\label{ward}

\vskip0.5cm
\noindent{\it The Ward identities of shift-symmetries.}\hskip0.3cm
The same four-fermion couplings $g_1$ and $g_2$ are assigned for both the left-
and right-handed sectors in the total action (\ref{stotal}). For the case 
of $a^2g_2\gg 1$ and $a^2g_1= 0$, the total
action (\ref{stotal}) possesses both the exact $\chi_R$-shift-symmetry 
(\ref{sshiftr}) and $\chi_L$-shift-symmetry (\ref{sshiftl}). 
These symmetries are not altered by the interactions between fermion 
fields and gauge fields of the RSM. As a consequence, the Ward identities 
associating to these symmetries completely determine all 1PI functions 
with external sterile neutrino fields $\chi_R$ and $\chi_L$. 

In the left-handed sector, the Ward identity associating to the
$\chi_R$-shift-symmetry is given by
($g_1\rightarrow 0$),
\begin{equation}
{1\over 2a}\gamma_\mu\partial^\mu\chi'_R(x)
+g_2\!\langle\Delta\!\left(\bar\psi_L(x)\!\cdot\!\left[
\Delta\chi_R(x)\right]\psi_L(x)\right)\rangle-{\delta\Gamma\over\delta\bar
\chi'_R(x)}=0,
\label{lw}
\end{equation}
where the ``primed'' fields are defined through the generating functional
approach, and ``$\Gamma$'' is the effective potential associating to the 
total action $S$ (\ref{stotal}) of the RSM (cf. section 2 in ref.\cite{xue97}). 
Analogously, in the right-handed sector, the Ward identity associating 
to the $\chi_L$-shift-symmetry is given by ($g_1\rightarrow 0$),
\begin{equation}
{1\over 2a}\gamma_\mu\partial^\mu\chi'_L(x)
+g_2\!\langle\Delta\!\left(\bar\psi_R(x)\!\cdot\!\left[
\Delta\chi_L(x)\right]\psi_R(x)\right)\rangle-{\delta\Gamma\over\delta\bar
\chi'_L(x)}=0.
\label{rw}
\end{equation}
The important consequences of these Ward identities can be seen 
in the following.

\noindent{\it Decoupling of left- and right- handed sectors.}\hskip0.3cm
Based on the Ward identities (\ref{lw}) for the left-handed sector and 
(\ref{rw}) for the right-handed sector, we can further take functional 
derivatives with respect to ``primed'' fields 
$\psi'_L,\chi'_R$ and $\psi'_R,\chi'_L$ respectively, and obtain all 1PI functions that 
contain external fields of both the left- and right-handed sectors:
\begin{equation}
{\delta^2\Gamma\over\delta\psi'_R\delta\bar
\chi'_R(x)}=0,\hskip0.2cm {\delta^2\Gamma\over\delta\chi'_L\delta\bar\chi'_R(x)}=0,
\hskip0.2cm \cdot\cdot\cdot,
\label{dlr}
\end{equation}
from the Ward identity (\ref{lw}) and
\begin{equation}
{\delta^2\Gamma\over\delta\psi'_L\delta\bar
\chi'_L(x)}=0,\hskip0.2cm {\delta^2\Gamma\over\delta\chi'_R\delta\bar
\chi'_L(x)}=0,\hskip0.2cm \cdot\cdot\cdot,
\label{drl}
\end{equation}
from the Ward identity (\ref{rw}). These Ward identities 
(\ref{dlr}) and (\ref{drl}) prove that all 1PI interacting vertices between the 
left- and right-handed sectors identically vanish. 
In addition, since the $\chi_L$- and $\chi_R$-shift-symmetries are 
exact and not altered by the gauge interactions of the RSM, the 
decoupling (\ref{dlr},\ref{drl}) between the left- 
and right-handed sectors is exact even in the presence of gauge fields 
$A_\mu$ of the RSM. This complete decoupling between the left- and right-handed 
sectors is so important that we separately deal with each sector.
 
\vskip0.5cm
\noindent{\it Decoupling of sterile neutrinos.}\hskip0.3cm
Taking functional derivative of the Ward identity (\ref{lw}) with 
respect to $\chi'_R$, we obtain\cite{xue97}
\begin{equation}
\int_x e^{-ipx} {\delta^{(2)}\Gamma\over\delta\chi'_R(x)\delta\bar\chi'_R(0)}
={i\over a}\gamma_\mu\sin(p^\mu)P_R\sim i\gamma_\mu\tilde p^\mu P_R,
\label{rfree}
\end{equation}
in the left-handed sector. Analogously, we obtain from the Ward identity (\ref{rw}),
\begin{equation}
\int_x e^{-ipx} {\delta^{(2)}\Gamma\over\delta\chi'_L(x)\delta\bar\chi'_L(0)}
={i\over a}\gamma_\mu\sin(p^\mu)P_L\sim i\gamma_\mu\tilde p^\mu P_L,
\label{lfree}
\end{equation}
in the right-handed sector. These 1PI functions show that the sterile 
neutrinos $\chi_R$ and $\chi_L$ do not receive any wave-function 
renormalizations.

Taking functional derivatives of the Ward identities eqs.(\ref{lw},\ref{rw}) with respect to 
the ``primed'' gauge and fermion fields: $A'_\mu$, $\psi'_L$ and $\psi'_R$, we obtain 
that all interacting 1PI functions with external gauge fields and sterile 
neutrino fields identically vanish: 
\begin{equation}
{\delta^2\Gamma\over\delta A'_\mu\delta\bar
\chi'_R(x)}=0,\hskip0.2cm {\delta^3\Gamma\over\delta A'_\mu
\delta\psi'_L\delta\bar\chi'_R(x)}=0,\hskip0.2cm \cdot\cdot\cdot,
\label{gld}
\end{equation}
for the left-handed sector and
\begin{equation}
{\delta^2\Gamma\over\delta A'_\mu\delta\bar
\chi'_L(x)}=0,\hskip0.2cm {\delta^3\Gamma\over\delta A'_\mu\delta\psi'_R\delta\bar
\chi'_L(x)}=0,\hskip0.2cm \cdot\cdot\cdot,
\label{grd}
\end{equation}
for the right-handed sector.
These results together with eqs.(\ref{rfree},\ref{lfree}) lead to the conclusion:
\begin{itemize}
\item{(i)}
two sterile neutrinos $\chi_R$ and $\chi_L$ completely decouple from gauge 
fields of the RSM;
\item{(ii)}
in the low-energy region ($p\sim 0$), the sterile neutrinos $\chi_R$ and $\chi_L$ 
are two free Weyl fermions.

\end{itemize}

\vskip0.5cm
\noindent{\it No hard spontaneous symmetry-breakings.}\hskip0.3cm
Further functional derivative of the Ward identity (\ref{lw}) with respect 
to $\psi'_L$ leads to the 1PI self-energy function $\Sigma^L(p)$ owing to the 
coupling between $\psi_L$ and $\chi_R$:
\begin{equation}
\int_x e^{-ipx} {\delta^{(2)}\Gamma\over\delta\psi'_L(x)\delta\bar\chi'_R(0)}
={1\over2}\Sigma^L(p)=0, \hskip0.5cm p=0,
\label{rws2}
\end{equation}
where identically vanishing $\Sigma^L(p)$ for $p=0$ is proved by eqs.(30) and 
(31) in ref.\cite{xue97}. In addition, the self-energy function $\Sigma^L(p)$ 
identically vanishes for $p\not=0$ as well:
\begin{equation}
\Sigma^L(p)=0\hskip0.5cm p\not=0,
\label{rws2'}
\end{equation}
which is proved by the strong coupling expansion $a^2g_2\gg 1$ (for details, see 
eq.(104) in \cite{xue97}). 
This shows no hard spontaneous breakings ($O(1/a)$) of chiral gauge symmetries (\ref{syl}), 
due to the non-vanishing v.e.v. $\langle\bar\psi_L\chi_R\rangle$, taking place in 
the left-handed sector.

Completely analogous to the analysis of the left-handed sector, further functional 
derivative of the Ward identity (\ref{rw}) with respect to $\psi'_R$ leads to the 
1PI self-energy function $\Sigma^R(p)$ owing to the coupling between $\psi_R$ and $\chi_L$:
\begin{equation}
\int_x e^{-ipx} {\delta^{(2)}\Gamma\over\delta\psi'_R(x)\delta\bar\chi'_L(0)}
={1\over2}\Sigma^R(p)=0,
\label{ws2}
\end{equation}
for both $p=0$ and $p\not=0$. This shows no hard spontaneous breakings of chiral 
gauge symmetries (\ref{syr}), due to the non-vanishing v.e.v. $\langle\bar\psi_R\chi_L\rangle$, 
taking place in the right-handed sector either.

These results are extremely crucial and necessary for:
\begin{itemize}

\item{(i)}
absolutely precluding any hard spontaneous symmetry-breakings in the both left-
and right-handed sectors of the RSM for $a^2g_1\rightarrow 0$ and $a^2g_2\gg 1$;

\item{(ii)}
possibly existing a low-energy scaling region of the RSM (\ref{stotal}) 
in a segment ${\cal A}$: $a^2g_1\rightarrow 0$ and $a^2g_2\gg 1$ in the phase 
space of four-fermion couplings $g_1$ and $g_2$.  

\end{itemize}

\vskip0.5cm
\noindent{\it 1PI vertices of four-fermion interactions.}\hskip0.3cm
In order to obtain the 1PI vertex-functions of the 
four-fermion interactions in both the left- and right-handed sectors, we respectively
take functional derivatives of the Ward identity (\ref{lw}) with 
respect to $\bar\psi'_L(0)$, $\psi'_L(y)$ and $\chi'_R(z)$, and the Ward identity
(\ref{rw}) with respect to $\bar\psi'_R(0)$, $\psi'_R(y)$ and $\chi'_L(z)$, and obtain:
\begin{eqnarray}
\int_{xyz}e^{-iyq-ixp-izp'}
{\delta^{(4)}\Gamma\over\delta\psi'_L(0)\delta\bar\psi'_L(y)\delta\chi'_R(z)
\delta\bar\chi'_R(x)}&=&4g_2w(p+{q\over 2})w(p'+{q\over 2}),
\label{4pl}\\
\int_{xyz}e^{-iyq-ixp-izp'}
{\delta^{(4)}\Gamma\over\delta\psi'_R(0)\delta\bar\psi'_R(y)\delta\chi'_L(z)
\delta\bar\chi'_L(x)}&=&4g_2w(p+{q\over 2})w(p'+{q\over 2}),
\label{4pr}
\end{eqnarray}
where $p+{q\over 2}$ and $p'+{q\over 2}$ are the external momenta of 
$\chi_R$($\chi_L$) fields; $p-{q\over 2}$ and $p'-{q\over 2}$ are the 
external momenta of $\psi_L$($\psi_R$) fields ($q$
is the energy-momentum transfer). 

These two identities eqs.(\ref{4pl},\ref{4pr}) show two consequences of the 
$\chi_R$- and $\chi_R$-shift-symmetries when $g_1=0$:

\begin{itemize}

\item{(i)}
the 1PI vertex-functions of four-fermion interactions are the 
exactly same as that at the tree-level without receiving any vertex-renormalization. 
This guarantees that the dimension-10 four-fermion interactions are only relevant 
operators for doublers in the high-energy region and irrelevant in the low-energy region 
(see the definition of the function $w(p)$ (\ref{swisf})). 

\item{(ii)}
large momentum states of $\chi_R(\chi_L)$ strongly couple to $\psi_L(\psi_R)$, 
while small momentum states of $\chi_R(\chi_L)$ weakly couple to $\psi_L(\psi_R)$.

\end{itemize}

\section{The strong- and weak-coupling analysis}\label{ana}

Aparting from the analysis based on the Ward identities associating to 
$\chi_{L,R}$-shift-symmetries, we use the strong- and weak-coupling 
expansions to analyze the RSM model. However, we do not present these 
analysis in details for the reasons that (i) the left- and right-handed 
sectors of the RSM are completely decoupled, as demonstrated in 
the last section (\ref{ward}); (ii) the four-fermion interactions (\ref{shil}) 
and (\ref{shir}) of both sectors are assigned with the same four-fermion couplings 
$g_1$ and $g_2$ and are the exactly same as that of the $SU_L(2)$ model and 
(11112) model, the only difference is in fermion contents; 
(iii) the strong- and weak-coupling analysis of the $SU_L(2)$ model 
and (11112) model are straightforwardly applied into the analysis of the 
left- and right-handed sectors of the RSM; (iv) 
the dynamics of the four-fermion interactions to realize a low-energy scaling 
region for the RSM in each 
sector is the same as that for the $SU_L(2)$ model and (11112) model; (v) 
the $SU_L(2)$ 
model was very intensively analyzed and the details of technique 
aspects are presented in refs.\cite{xue97,xue99,xue00}. 
In the following sections, we will indicate the main aspects of techniques 
adopted by giving clear references to the counterparts in the $SU_L(2)$ model
in refs.\cite{xue97,xue99,xue00}.  

The RSM action for analyzing the spectrum and 1PI functions in the case 
of the $a^2g_1\rightarrow 0 $ and strong coupling $a^2g_2\gg 1$ is obtained by 
rescaling all fermion fields $\psi\rightarrow g_2^{1\over4}\psi$ to be 
dimensionless. As a result, we rewrite the RSM total action as,
\begin{eqnarray}
S&=&{1\over 2ag_2^{1\over2}}\sum_x\Big(\bar\psi_L(x) D^L_\mu
\cdot\gamma_\mu\psi_L(x)
+\bar\psi_L(x) D^L_\mu\cdot\gamma_\mu\psi_L(x)+\cdot\cdot\cdot\Big)\nonumber\\
 &+& \sum_x \left(S^L_i(x)+ S^R_i(x)\right),
\label{saction}
\end{eqnarray}
where ``$\cdot\cdot\cdot$" stands for the kinetic terms (\ref{sterile}) 
for two sterile neutrinos $\chi_L$ and $\chi_R$, the coupling $g_2$ of 
the dimension-10 operators in $S^{L,R}_i(x)$ is rescaled away and the 
dimension-6 operators ($g_1$) in $S^{L,R}_i(x)$ are disregarded. 
The dimension-10 operators are only relevant in the 
high-energy region, where non-vanishing value of the operator $\omega(p)\not=0$ 
(\ref{swisf}) for $p_\mu=\pi_A$ gives rise to the non-trivial strong coupling 
limit (for details, see the appendix in ref.\cite{xue97}).

\section{Massive fermion states in the RSM}\label{fermion}

For strong four-fermion interactions (\ref{shil}) and (\ref{shir}), i.e., the 1PI vertices 
(\ref{4pl}) and (\ref{4pr}) for $p,p'\sim \pi_A$ and $a^2g_2\gg 1$, three-fermion
states are formed. The three-fermion states in the RSM can be classified into 
three categories:
(i) the neutral sector in both left- and right-handed sectors, (ii) the charge 
sector in the left-handed sector, (iii) the charged sector in the right-handed
sector. We first discuss the neutral sector, in which fermion propagators, mass terms 
$M(p)$ and form factor $Z(p)$ of neutral three-fermion states can be exactly 
obtained by using Ward identities (\ref{lw}) and (\ref{rw}).

\vskip0.5cm
\noindent{\it Neutral sectors.}\hskip0.3cm
The neutral three-fermion singlets $\Psi^n_L$ and $\Psi^n_R$
in the left- and right-handed sector of the RSM are: 
\begin{eqnarray}
\Psi^n_L&\equiv& {1\over2a}(\bar\psi_L\cdot\chi_R)\psi_L,\hskip 0.3cm 
\psi_L=\left(\matrix{\nu\cr e}\right)_L,\left(\matrix{u\cr d}\right)_L;\cdot\cdot\cdot,
\label{nl3}\\
\Psi^n_R&\equiv& {1\over2a}(\bar\psi_R\cdot\chi_L)\psi_R,\hskip 0.3cm 
\psi_R=e_R, u_R, d_R;\cdot\cdot\cdot,
\label{nr3}
\end{eqnarray}
both are Weyl fermions.
The singlets $\Psi^n_L$ ($\Psi^n_R$) are neutral as the sterile 
neutrino $\chi_R$ ($\chi_L$), but opposite chiralities. The three-fermion singlets 
$\Psi^n_L$ and $\Psi^n_R$ respectively combine together with sterile neutrinos $\chi_R$  
and $\chi_L$ to form neutral massive Dirac fermions:
\begin{eqnarray}
\Psi_n^L= (\Psi^n_L|_{\rm ren}, \chi_R),\hskip0.3cm \Psi^n_L|_{\rm ren}&=&Z^{-1}_n(\pi_A)\Psi^n_L,
\label{r4ln1}\\
\Psi_n^R= (\Psi^n_R|_{\rm ren}, \chi_L),\hskip0.2cm \Psi^n_R|_{\rm ren}&=&Z^{-1}_n(\pi_A)\Psi^n_L,
\label{r4ln2}
\end{eqnarray}
consistently with the global symmetries $U_{\chi_L}(1)$, $U_{\chi_R}(1)$ and 
$U_{L,R}(1)$ of the RSM. $\Psi^n_{L,R}|_{\rm ren}$ in eqs.(\ref{r4ln1}, \ref{r4ln2}) are the 
renormalized three-fermion singlets at $p\simeq\pi_A$. By using the Ward identities 
(\ref{lw}) and (\ref{rw}), we obtain the masses of these Dirac fermions $\Psi_n^L$ 
and $\Psi_n^R$
\begin{equation}
{1\over2}M(p)=\int_xe^{-ipx}
{\delta^{(2)}\Gamma\over\delta\Psi'^n_R(x)\delta\bar\chi'_L(0)}=\int_xe^{-ipx}
{\delta^{(2)}\Gamma\over\delta\Psi'^n_L(x)\delta\bar\chi'_R(0)}=8ag_2w^2(p),
\label{fm}
\end{equation}
and the form factors $Z_n(p)$ of the three-fermion
singlets $\Psi^n_L$ and $\Psi^n_R$,
\begin{equation}
Z_n(p)=a\int_xe^{-ipx}
{\delta^{(2)}\Gamma\over\delta\Psi'^n_R(x)\delta\bar\chi'_L(0)}=a\int_xe^{-ipx}
{\delta^{(2)}\Gamma\over\delta\Psi'^n_L(x)\delta\bar\chi'_R(0)}=aM(p).
\label{zn}
\end{equation}
In the high-energy region $p\sim \pi_A$, both the mass $M(p)$ and form factor 
$Z_n(p)$ are:
\begin{equation}
M(\pi_A)={\rm const.},\hskip0.5cm Z_n(\pi_A)={\rm const.}.
\label{mz}
\end{equation}
  
In the high-energy region $p=\tilde pa +\pi_A$, the propagators of these 
neutral and massive Dirac fermions are obtained by using the Ward identities 
(\ref{lw}) and (\ref{rw}), 
\begin{equation}
S^{R,L}_n(p)=\int_xe^{-ipx}\langle\Psi^{R,L}_n(0)\bar\Psi^{R,L}_n(x)\rangle=
{{i\over a}\gamma_\mu \sin (p)^\mu +M(p)
\over {1\over a^2}\sin^2p + M^2(p)}\simeq {i\gamma_\mu \tilde p^\mu +{1\over a}
\over \tilde p^2 + ({1\over a})^2}.
\label{dnp'}
\end{equation}
The neutral mass terms are given by
\begin{equation}
\bar\Psi^n_L|_{\rm ren}\chi_R + \bar\chi_R\Psi^n_L|_{\rm ren},
\hskip0.5cm \bar\Psi^n_R|_{\rm ren}\chi_L 
+ \bar\chi_L\Psi^n_R|_{\rm ren},
\label{mr4ln}
\end{equation}
at the mass scale $1/a$.

\vskip0.5cm
\noindent{\it Charged sectors.}\hskip0.3cm
The three-fermion doublets $\Psi_R$ and singlets $\Psi_L$ are: 
\begin{eqnarray}
\Psi_R&\equiv& {1\over2a}(\bar\chi_R\cdot\psi_L)\chi_R,\hskip 0.3cm 
\psi_L=\left(\matrix{\nu\cr e}\right)_L,\left(\matrix{u\cr d}\right)_L;\cdot\cdot\cdot,
\label{l3}\\
\Psi_L&\equiv& {1\over2a}(\bar\chi_L\cdot \psi_R)\chi_L,\hskip 0.5cm 
\psi_R=e_R, u_R, d_R;\cdot\cdot\cdot,
\label{r3}
\end{eqnarray}
in the left- and right-handed sectors of the RSM. They are Weyl fermions respectively
having the same gauge charges as the left-handed 
doublets $\psi_L$ and right-handed singlets $\psi_R$, but opposite chiralities. 
The three-fermion doublets $\Psi_R$ and singlets $\Psi_L$ respectively combine 
together with $\psi_L$ and $\psi_R$ to form massive Dirac fermions:
\begin{eqnarray}
\Psi^L_c&=&(\psi_L, \Psi_R|_{\rm ren}),\hskip0.3cm\Psi_R|_{\rm ren}=Z^{-1}_L(\pi_A)\Psi_R,
\label{r4l}\\
\Psi^R_c&=&(\psi_R, \Psi_L|_{\rm ren}),\hskip0.3cm \Psi_L|_{\rm ren}=Z^{-1}_R(\pi_A)\Psi_L.
\label{r4r}
\end{eqnarray}
$\Psi^L_c$ is consistent with the gauge symmetries $SU^L_c(3)\otimes SU_L(2)\otimes U^L_Y(1)$ 
of the left-handed sector and $\Psi^R_c$ is consistent with $SU^R_c(3)\otimes U^R_Y(1)$ of the 
right-handed sector. Note that $SU^{L,R}_c(3)$ is only for the quark sector. $\Psi_{R,L}|_{\rm ren}$ in eqs.(\ref{r4l},\ref{r4r}) are the 
renormalized charged three-fermion doublets and singlets at $p\simeq\pi_A$. The form factors $Z_{L,R}(p)$ 
of the three-fermion states 
$\Psi_{R,L}$ are respectively computed by the strong coupling expansion 
for $p\simeq\pi_A$\cite{xue99}
\begin{eqnarray}
Z_L(p)&=&a\int_xe^{-ipx}
{\delta^{(2)}\Gamma\over\delta\Psi'_R(x)\delta\bar\psi'_L(0)}\simeq aM(p),
\label{zl}\\
Z_R(p)&=&a\int_xe^{-ipx}
{\delta^{(2)}\Gamma\over\delta\Psi'_L(x)\delta\bar\psi'_R(0)}\simeq aM(p).
\label{zr}
\end{eqnarray}
In the high-energy region $p=\tilde pa +\pi_A$, the propagators of these 
charged and massive Dirac fermions are obtained by the strong coupling expansion,
\begin{eqnarray}
S^L_c(p)&=&\int_xe^{-ipx}\langle\Psi^L_c(0)\bar\Psi^L_c(x)\rangle \simeq
{{i\over a}\gamma_\mu \sin (p)^\mu +M(p)
\over {1\over a^2}\sin^2p + M^2(p)}\delta^L\simeq {i\gamma_\mu \tilde p^\mu +{1\over a}
\over \tilde p^2 + ({1\over a})^2}\delta^L,
\label{dlp'}\\
S^R_c(p)&=&\int_xe^{-ipx}\langle\Psi^R_c(0)\bar\Psi^R_c(x)\rangle \simeq
{{i\over a}\gamma_\mu \sin (p)^\mu +M(p)
\over {1\over a^2}\sin^2p + M^2(p)}\delta^R\simeq {i\gamma_\mu \tilde p^\mu +{1\over a}
\over \tilde p^2 + ({1\over a})^2}\delta^R,
\label{drp'}
\end{eqnarray} 
where $\delta^{L,R}$ are the identity matrices respectively associating to the gauge groups in 
the left- and right-handed sectors of the RSM. The gauge-invariant mass term is given by
\begin{equation}
\bar\psi_L\Psi_R|_{\rm ren} + \bar\Psi_R|_{\rm ren}\psi_L,
\hskip0.3cm
\bar\psi_R\Psi_L|_{\rm ren} + \bar\Psi_L|_{\rm ren}\psi_R,
\label{mr4r}
\end{equation}
at the mass scale $1/a$.

All these three-fermion states (with the dimension [$a^{1/2}$]) and massive Dirac 
fermions are formed in the high-energy region 
$p_\mu\simeq \pi_A$, extra fermion species (doublers) at 
$p_\mu\simeq \pi_A$ are thus decoupled from the low-energy spectrum. Note that these
massive Dirac fermions can be degenerate, provided we properly make the wave-function
renormalization for each fermion (doubler) at $p\simeq \pi_A$ in the edges of 
the Brillouin zone. 
The three-fermion doublets (\ref{l3}) and singlets (\ref{r3}) are not only the 
eigen-states of the gauge invariant mass $M(p)$, but also 
eigen-states of gauge interactions of the RSM so that their mass terms (\ref{mr4r}) 
preserve chiral gauge symmetries of the RSM.

\section{Massive boson states in the RSM}\label{boson}

For strong four-fermion interactions (\ref{shil}) and (\ref{shir}), i.e., 1PI vertices 
(\ref{4pl}) and (\ref{4pr}) for $p,p'\sim \pi_A$, $a^2g_1\rightarrow 0$ 
and $a^2g_2\gg 1$, two types 
of massive composite scalar fields $\Phi_L$ and $\Phi_R$ are formed. In this section,
we discuss the charges and masses of these composite scalar fields:

For the left-handed sector,  
\begin{eqnarray}
\Phi_L&=&\bar\chi_R\cdot\psi_L,\hskip 0.3cm 
\psi_L=\left(\matrix{\nu\cr e}\right)_L,\left(\matrix{u\cr d}\right)_L;\cdot\cdot\cdot,
\label{lcomps}\\
\Phi_R&=&\bar\chi_L\cdot\psi_R,\hskip 0.3cm 
\psi_R=e_R, u_R, d_R;\cdot\cdot\cdot.
\label{rcomps}
\end{eqnarray}
The real and imaginary parts of $\Phi_L$ are four composite scalars
\begin{eqnarray}
\Phi_L^1&=&{1\over2}(\bar\psi_L\cdot\chi_R+\bar\chi_R\cdot\psi_L)
\nonumber\\
\Phi_L^2&=&{i\over2}(\bar\psi_L\cdot\chi_R-\bar\chi_R\cdot\psi_L).
\label{lreals}
\end{eqnarray}
The real and imaginary parts of $\Phi_R$ are two composite scalars
\begin{eqnarray}
\Phi_R^1&=&{1\over2}(\bar\psi_R\cdot\chi_L+\bar\chi_L\cdot\psi_R)
\nonumber\\
\Phi_R^2&=&{i\over2}(\bar\psi_R\cdot\chi_L-\bar\chi_L\cdot\psi_R).
\label{rreals}
\end{eqnarray}

These boson states $\Phi_L$ and $\Phi_R$ respectively have the same gauge 
charges as that of $\psi_L$ and $\psi_R$, 
but carry the charges of global symmetries $U_{\chi_R}(1)$ and $U_{\chi_L}(1)$. 
By the strong coupling expansion, analogously to the analysis given in ref.\cite{xue97}, 
we compute these composite scalars and their propagators:
\begin{eqnarray}
G^L(q)&=&\int_x e^{-iqx}\langle\Phi_L(0),\Phi^\dagger_L(x)\rangle,
\label{lbosonp}\\
G^R(q)&=&\int_x e^{-iqx}\langle\Phi_R(0),\Phi^\dagger_R(x)\rangle,
\label{rbosonp}
\end{eqnarray}
where $q=p'-p$ is the energy-momentum transfer and $p,p'\sim \pi_A, q\sim 0$.
Making wave-function renormalization,
\begin{eqnarray}
\Phi^r_L&=&(Z_L^b)^{-1}(\pi_A)\Phi_L,\hskip 0.3cm Z_L^b(p)=2w(p),
\label{lrenb}\\
\Phi^r_R&=&(Z_R^b)^{-1}(\pi_A)\Phi_R,\hskip 0.3cm Z_R^b(p)=2w(p),
\label{rrenb}
\end{eqnarray}
we find the propagators for these massive composite scalar modes,
\begin{eqnarray}
G^L(q)&=& {\delta_L\over {4\over a^2}\sum_\mu\sin^2{q_\mu\over 2}
+\mu^2};\label{lscalar}\\
G^R(q)&=& {\delta_R\over {4\over a^2}\sum_\mu\sin^2{q_\mu\over 2}
+\mu^2};\label{rscalar}\\
\mu^2 &=& 4(g_2-{2\over a^2})>0,
\label{bmas}
\end{eqnarray}
which $\delta_{L,R}$ are factors stemming from the gauge group of the 
left- and right-handed sectors. 
The $Z_L^b(p)$ and $Z_R^b(p)$ are form factors of the scalar fields $\Phi_L$ 
and $\Phi_R$. They are constants for $p\simeq\pi_A$ and vanishing for $p\sim 0$. 

The positive mass terms in these boson propagators ($\mu^2>0$) indicate that the 
1PI functions relating to gauge invariant operators $\Phi^r_L\Phi^{r\dagger}_L$
(dimension-2), $\Phi^r_L \Phi^{r\dagger}_L\Phi^r_L\Phi^{r\dagger}_L$ (dimension-4)
in the left-handed sector and the counterparts $\Phi^r_R\Phi^{r\dagger}_R$,
$\Phi^r_R \Phi^{r\dagger}_R\Phi^r_R\Phi^{r\dagger}_R$ in the right-handed sector
are definitely positive. This means that $a^2g_2\gg 1$ is a symmetric phase, no
spontaneous symmetry-breaking and the energy of ground states of the RSM is bound 
from the bellow.

\section{Vectorlike gauge coupling vertices in the RSM}\label{gauge}

The gauge symmetries 
$SU^L_c(3)\otimes SU_L(2)\otimes U^L_Y(1)$ in the left-handed sector and 
$SU^R_c(3)\otimes U^R_Y(1)$ in the right-handed sector of the RSM are respectively realized 
by the vectorlike Dirac spectra (\ref{r4ln1}), \ref{r4ln2}), (\ref{r4l}) and (\ref{r4r}) in 
the high-energy region. In this section, we 
turn on gauge fields of the RSM as dynamical fields to examine the
1PI interacting vertices of gauge and fermion fields. The 1PI functions of gauge-fermion
couplings in general are made by all possible gauge invariant operators comprising by 
Dirac fields (e.g.,(\ref{r4l})) and gauge links $U_\mu$. However, 
low-dimension relevant operators are important for perturbative expansion 
$U_\mu\simeq 1+iaA_\mu+\cdot\cdot\cdot$. 

We first discuss such 1PI vertex
functions in the neutral sectors of the RSM, since they can be exactly obtained by the 
Ward identities (\ref{lw}) and (\ref{rw}). Then, we discuss the 1PI 
vertex functions of gauge-fermion couplings in the left- and right-handed sectors of 
the RSM. Both of them can be respectively obtained by the Ward identities associating 
to gauge symmetries and the strong coupling expansions. 

\vskip0.5cm
\noindent{\bf \it Decoupling of the neutral sectors.}\hskip0.3cm
We consider all possible 1PI interacting vertices
involving external gauge field $A'_\mu$ and fermion fields $\chi'_R, \Psi'^n_R$ 
and $\chi'_L, \Psi'^n_L$ in the neutral sectors of the RSM. 
Based on the Ward identities (\ref{lw}) and (\ref{rw}) 
of the $\chi_R$- and $\chi_L$-shift-symmetries, 
we take functional derivatives with respect to the external gauge field $A'_\mu$, 
fermion fields $\chi'_R,\chi'_L$ and $\Psi'^n_L,\Psi'^n_L$, to obtain the following Ward 
identities: 
\begin{equation}
{\delta^{(2)} \Gamma\over\delta
A'_\mu\delta\bar\chi'_R}={\delta^{(3)} \Gamma\over\delta
A'_\mu\delta\chi'_R\delta\bar\psi'_L}={\delta^{(3)} \Gamma\over\delta
A'_\mu\delta\Psi'^n_R\delta\bar\chi'_R}=\cdot\cdot\cdot=0,
\label{wanl}
\end{equation}
for the left-handed sector, and
\begin{equation}
{\delta^{(2)} \Gamma\over\delta
A'_\mu\delta\bar\chi'_L}={\delta^{(3)} \Gamma\over\delta
A'_\mu\delta\chi'_L\delta\bar\psi'_R}={\delta^{(3)} \Gamma\over\delta
A'_\mu\delta\Psi'^n_L\delta\bar\chi'_L}=\cdot\cdot\cdot=0,
\label{wa}
\end{equation}
for the right-handed sector. 
These Ward identities show the identical vanishing of all interacting 
1PI functions containing external gauge field $A'_\mu$, fermion fields $\chi'_R, \Psi'^n_R$ 
in the left-handed sector and $\chi'_L, \Psi'^n_L$ in the right-handed sector. 
This indicats no any interactions
between the gauge field $A_\mu$ and neutral fermions $\chi_R, \Psi^n_R;
\chi_L, \Psi^n_L$. It should be emphasized that this decoupling is valid not
only for perturbative gauge-interaction but also non-perturbative gauge-interaction, 
since the RSM action (\ref{stotal}) is exactly chiral-gauge symmetric and 
$\chi_L$- and $\chi_R$-shift-symmetric for any values of gauge couplings.

As a result, the left- and right-handed neutral currents $j^\mu_R$ and $j^\mu_L$ 
associating to the global symmetries $U_{\chi_R}(1)$ and $U_{\chi_L}(1)$ for 
two sterile neutrinos $\chi_R$ and $\chi_L$:
\begin{eqnarray}
j^\mu_R&=&i\bar\chi_R\gamma^\mu\chi_R,\hskip0.3cm\partial_\mu j_R^\mu =0;\label{nlcc}\\ 
j^\mu_L&=&i\bar\chi_L\gamma^\mu\chi_L,\hskip0.3cm\partial_\mu j_L^\mu =0;\label{nrcc}
\end{eqnarray}
are exactly conserved and no anomalous contribution is expected from the topological gauge 
configuration. This is a very important feature, which we will see later in 
section (\ref{sanomaly}), for obtaining the gauge anomaly and fermion-flavour singlet 
anomalies relating to violations of the $U_L(1)$-symmetry (the number of the fermion field 
$\psi_L$ in the left-handed sector) and $U_R(1)$-symmetry (the number of the fermion field 
$\psi_R$ in the right-handed sector).

\vskip0.5cm
\noindent{\it Vectorlike gauge couplings.}\hskip0.3cm
Since the massive Dirac fermions (\ref{r4l}) and (\ref{r4r}) are vectorially gauged,
all 1PI functions of gauge-fermion couplings must be vectorlike. 
In order to reveal the vectorlike feature of the 1PI functions of gauge-fermion couplings, 
in terms of perturbative gauge fields $A_\mu$ order by order, we consider the
1PI functions of Dirac fermions (\ref{r4l}) and (\ref{r4r}) coupling to $A_\mu$.
 
In the left-handed sector, the interacting 1PI vertex function between the perturbative
gauge fields $A_\mu$ and renormalized charged Dirac fermion $\Psi^L_c$ 
(\ref{r4l}) is related to the following three-point Green functions,
\begin{eqnarray}
\langle\Psi^L_c(x_1)\bar\Psi^L_c(x) A_\nu(y)\rangle
&=&\langle\psi_L(x_1)\bar\psi_L(x) A_\nu(y)\rangle
+\langle\psi_L(x_1)\bar\Psi_R(x)|_{\rm ren} A_\nu(y)\rangle
\nonumber\\
&+&\langle\Psi_R(x_1)|_{\rm ren}\bar\psi_L(x) A_\nu(y)\rangle
+\langle\Psi_R(x_1)|_{\rm ren}\bar\Psi_R(x)|_{\rm ren} A_\nu(y)\rangle,
\label{l3points}
\end{eqnarray}
and
\begin{equation} 
\int_{x_1xy}e^{i(p'x\!+\!px_1\!-\!qy)}\langle\Psi^L_c(x_1)\bar\Psi^L_c(x) A_\nu(y)\rangle
\!=\! G_{\nu\mu}(q)S^L_c(p) \Lambda^L_{\mu c}(p,p',\Psi_R|_{\rm ren})S^L_c(p'),
\label{lpad}
\end{equation}
where $p,p'\sim \pi_A$ are the external momenta of fermion fields and
transfer momentum $q=p-p'$. 
The $\Lambda^L_{\mu c}(p,p',\Psi_R|_{\rm ren})$ is the renormalized 1PI vertex function 
of the gauge-fermion couplings in the left-handed sector. $S^L_c(p)$ is the 
propagator (\ref{dlp'}) of the renormalized charged Dirac fermion $\Psi^L_c$ (\ref{r4l}). 
From eq.(\ref{l3points}), this renormalized 1PI vertex function can be written as,
\begin{eqnarray}
\Lambda^L_{\mu c}(p,p',\Psi_R|_{\rm ren})&=&\Lambda^L_{\mu LL}(p,p')
+\Lambda^L_{\mu LR}(p,p',\Psi_R|_{\rm ren})\nonumber\\
&+&
\Lambda^L_{\mu RL}(p,p',\Psi_R|_{\rm ren})+\Lambda^L_{\mu RR}(p,p',\Psi_R|_{\rm ren}).
\label{lvdirac}
\end{eqnarray}

In the right-handed sector, the interacting 1PI vertex function between the gauge 
fields $A_\mu$ and renormalized charged Dirac fermion $\Psi^R_c$ 
(\ref{r4r}) is related to the following three-point Green functions,
\begin{eqnarray}
\langle\Psi^R_c(x_1)\bar\Psi^R_c(x) A_\nu(y)\rangle
&=&\langle\psi_R(x_1)\bar\psi_R(x) A_\nu(y)\rangle
+\langle\psi_R(x_1)\bar\Psi_L(x)|_{\rm ren} A_\nu(y)\rangle
\nonumber\\
&+&\langle\Psi_L(x_1)|_{\rm ren}\bar\psi_R(x) A_\nu(y)\rangle
+\langle\Psi_L(x_1)|_{\rm ren}\bar\Psi_L(x)|_{\rm ren} A_\nu(y)\rangle,
\label{r3points}
\end{eqnarray}
and
\begin{equation} 
\int_{x_1xy}e^{i(p'x\!+\!px_1\!-\!qy)}\langle\Psi^R_c(x_1)\bar\Psi^R_c(x) A_\nu(y)\rangle
\!=\! G_{\nu\mu}(q)S^R_c(p) \Lambda^R_{\mu c}(p,p',\Psi_L|_{\rm ren})S^R_c(p'),
\label{rpad}
\end{equation}
where $p,p'\sim \pi_A$ are the external momenta of fermion fields and
transfer momentum $q=p-p'$. 
The $\Lambda^R_{\mu c}(p,p',\Psi_L|_{\rm ren})$ is the renormalized 1PI vertex function 
of the gauge-fermion couplings in the right-handed sector. $S^R_c(p)$ is the 
propagator (\ref{drp'}) of the renormalized 
charged Dirac fermion $\Psi^R_c$ (\ref{r4r}). From eq.(\ref{r3points}), this 
the renormalized 1PI vertex function can be written as,
\begin{eqnarray}
\Lambda^R_{\mu c}(p,p')&=&\Lambda^R_{\mu RR}(p,p')+\Lambda^R_{\mu RL}(p,p',\Psi_L|_{\rm ren})
\nonumber\\
&+&
\Lambda^R_{\mu LR}(p,p',\Psi_L|_{\rm ren})+\Lambda^R_{\mu LL}(p,p',\Psi_L|_{\rm ren}).
\label{rvdirac}
\end{eqnarray}

Each term in the 1PI vertex functions (\ref{lvdirac})) and (\ref{rvdirac}) can be explicitly 
computed by the strong coupling expansion and perturbative expansion of small gauge 
couplings order by order\cite{xue00}. 
The 1PI vertex functions 
(\ref{lvdirac}) and (\ref{rvdirac})  can be divided into two parts: the first part 
without external three-fermion doublets(singlets) $\Psi_R|_{\rm ren}$($\Psi_L|_{\rm ren}$); 
the second part with external three-fermion doublets(singlets) $\Psi_R|_{\rm ren}$
($\Psi_L|_{\rm ren}$). In eqs.(\ref{lvdirac}) and (\ref{rvdirac}), the 
second and third terms, $p'$ is the external momentum of three-fermion doublets(singlets), 
while in the last term, both $p'$ and $p$ are the external momenta 
of three-fermion doublets(singlets). 

Ananlogously to the analysis in ref.\cite{xue00}), we can show that the renormalized 1PI vertex 
functions (\ref{lvdirac}) and (\ref{rvdirac}) precisely obey the following Ward 
identities of the exact chiral gauge symmetries of the RSM at the lattice scale $1/a$.
The Ward identity associating to the $SU^L_c(3)\otimes SU_L(2)\otimes U^L_Y(1)$ 
gauge symmetries of the left-handed sector is given by,
\begin{equation}
\left({i\over a}\right)(
\sin p_\mu-\sin p'_\mu)\Lambda^L_{\mu c}(p,p',\Psi_R|_{\rm ren})=\left(S^L_c(p)\right)^{-1}
-\left(S^L_c(p')\right)^{-1}.
\label{gwardl}
\end{equation}
The Ward identity associating to the $SU^L_c(3)\otimes U^R_Y(1)$ gauge symmetries
of the right-handed sector is given by,
\begin{equation}
\left({i\over a}\right)(
\sin p_\mu-\sin p'_\mu)\Lambda^R_{\mu c}(p,p',\Psi_L|_{\rm ren})=\left(S^R_c(p)\right)^{-1}
-\left(S^R_c(p')\right)^{-1}.
\label{gwardr}
\end{equation}
In eqs.(\ref{gwardl}) and (\ref{gwardr}), the gauge coupling and gauge group generators 
are eliminated from the vertex functions $\Lambda^L_{\mu c}(p,p',\Psi_R|_{\rm ren})$ and 
$\Lambda^R_{\mu c}(p,p',\Psi_L|_{\rm ren})$. Eqs.(\ref{gwardl}) and (\ref{gwardr}) show 
that the exact chiral gauge 
symmetries are preserved by the vectorlike and massive spectra of Dirac fermions 
(\ref{r4l}) and (\ref{r4r}) in the RSM. 

\vskip0.5cm
\noindent{\it Gauge coupling to boson states.}\hskip0.3cm
Massive boson states $\Phi^r_L$ couple to the gauge fields of 
$SU^L_c(3)\otimes SU_L(2)\otimes U^L_Y(1)$ in the left-handed sector.  
Massive boson states $\Phi^r_R$ couple to the gauge fields of 
$SU^L_c(3)\otimes U^L_Y(1)$ in the right-handed sector. the 1PI functions 
of the gauge-boson couplings can be constructed by 
gauge invariant operators comprising field operators $\Phi^r_L$ and $U_\mu$ 
($\Phi^r_R$ and $U_\mu$), analogously to the 1PI functions
of elementary scalar fields coupling to gauge fields.

\section{The low-energy scaling region for the RSM}\label{declr}

In previous sections, we have discussed  the spectra of fermions and bosons and 
their gauge-couplings for $p\sim \pi_A$ 
and $a^2g_2\gg 1$. In this section, we turn 
to discussions of the spectra and gauge-couplings in the low-energy region where 
$p=a\tilde p\rightarrow 0$ and $a^2g_2\gg 1$. We first briefly recall the 
the low-energy scaling region advocated in the $SU_L(2)$ model\cite{xue97}.

\vskip0.5cm
\noindent{\it The low-energy scaling region.}\hskip0.3cm
In the $SU_L(2)$ model, it is shown there exists a low-energy 
scaling region in a peculiar segment in the phase space of the four-fermion 
couplings $g_1,g_2$: 
\begin{equation}
{\cal A}=\Big[g_1\rightarrow 0, g_2^{c,a}<g_2<g_2^{c,\infty}\Big],\hskip0.3cm
a^2g_2^{c,a}=0.124,\hskip0.3cm
1\ll g_2^{c,\infty}< \infty, 
\label{segment}
\end{equation}
for chiral $SU_L(2)$-gauged fermions in the low-energy limit $p=a\tilde p\rightarrow 0$. 
$g_2^{c,\infty}$ is a finite number and $g_2^{c,a}$ indicates the
critical value above which the effective four-fermion couplings relevant to all 
doublers ($p\sim \pi_A$) are strong enough, so that all doublers are gauge-invariantly 
decoupled, as very massive Dirac fermions. 
In refs.\cite{xue99} and \cite{xue00}, by using Ward identities, strong and weak
couplings expansions, as well as the analytical properties of 1PI functions in the energy-momentum space, 
we discuss the phenomenon of three-fermion states 
dissolving into corresponding three-fermion cuts, because three-fermion state's 
mass-gaps and form factors vanish in this low-energy scaling region. 

The low-energy scaling region of the RSM must be the qualitatively 
same as the low-energy scaling region ${\cal A}$ 
(\ref{segment}) of the $SU_L(2)$ model for the reasons that 
\begin{itemize}

\item{(i)} the left- and right-handed sectors of the RSM are completely decoupled;

\item{(ii)} both sectors have the same type and strength of four-fermion couplings $g_2$ as 
that of the $SU_L(2)$ model. The gauge symmetries and global symmetries of each sector are 
analogous to that of the $SU_L(2)$ model. The only difference is just fermion contents, 
which must not qualitatively change dynamics of realizing the low-energy scaling 
region ${\cal A}$ 
(\ref{segment}).

\end{itemize} 
We first stress the crucial points for the dynamics of the four-fermion interactions 
(\ref{shil}) and (\ref{shir}) of the RSM to realize this low-energy scaling 
region ${\cal A}$: 
(i) 1PI functions for four-fermion interactions (\ref{4pl}) and (\ref{4pr}) do not 
receive any vertex-function renormalizations and vanish as the energy-momenta of 
the sterile neutrinos 
$\chi_R$ and $\chi_L$ go to zero; (ii) no hard spontaneous symmetry-breakings take place
(\ref{rws2},\ref{rws2'},\ref{ws2}); (iii) the mass-gaps and form factors of three-fermion 
doublets $\Psi_R|_{\rm ren}$, singlets $\Psi_L|_{\rm ren}$ and neutral states 
$\Psi^n_{R,L}|_{\rm ren}$ vanish 
in the low-energy scaling region.

\vskip0.5cm
\noindent{\it The dissolving phenomenon.}\hskip0.3cm
The three-fermion states: neutral singlets $\Psi^n_{R,L}|_{\rm ren}$ 
(\ref{r4ln1},\ref{r4ln2}), charged doublets $\Psi_R|_{\rm ren}$ (\ref{r4l})
and charged singlets $\Psi_L|_{\rm ren}$ (\ref{r4r}), 
are characterized by their binding energies $M(p)$'s (mass-gaps) and form factors $Z(p)$'s.
$M(p)$'s and $Z(p)$'s are respectively related to the poles and residues of the composite
Dirac fermion propagators (\ref{dnp'}), (\ref{dlp'}) and (\ref{drp'}) (see section V). 
They are obtained by the Ward 
identities (\ref{lw},\ref{rw}) and strong coupling expansion for $p\simeq \pi_A$ and 
$a^2g_2\gg 1$. The explicit expressions for $M(p)$'s and $Z(p)$'s are given in 
eqs.(\ref{fm},\ref{zn}) and (\ref{zl},\ref{zr}), which are energy-momentum dependent. Since 
the total action (\ref{stotal}) of the RSM is local, all 1PI functions are continuous 
and analytical functions in the energy-momentum 
space. Given a fixed value of the four-fermion couplings $a^2g_2\gg 1$, we can analytically 
continue $M(p)$'s and $Z(p)$'s that obtained in the high-energy 
region $p\simeq\pi_A$ to the low-energy region $p\rightarrow 0$. For $p\sim 0$, we have,
\begin{equation}
M(p)'s\rightarrow O(p^4),\hskip0.5cm Z(p)'s\rightarrow O(p^4),
\label{mz0}
\end{equation}
which are on the other hand verified by the weak coupling expansion 
for $p\rightarrow 0$ and $a^2g_2\gg 1$\cite{xue99}. 

Due to the vanishing of form factors $Z(p)$'s for $p\sim 0$ (\ref{mz0}), we are not allowed
to make the wave-function renormalizations in eqs.(\ref{r4ln1},\ref{r4ln2},\ref{r4l}) 
and (\ref{r4r}). The Dirac fermion propagators (\ref{dnp'}), (\ref{dlp'}) 
and (\ref{drp'}) for the neutral three-fermion states 
$\Psi^n_{R,L}$ and the charged three-fermion states $\Psi_R,\Psi_L$ are no longer valid for 
$p\rightarrow 0$ and thus do not represent massless poles at $p=0$. In fact, the 
two-point Green functions of the composite operators $\Psi^n_{R,L}$, $\Psi_R$ and $\Psi_L$ go 
to zero, because their residues go to zero when $p\rightarrow 0$ (\ref{mz0}). This means that as discussed in the (11112) 
model and $SU_L(2)$ model in refs.\cite{xue99,xue00,11112}, 
the three-fermion states: the doublets $\Psi_L$ and singlets $\Psi_R$ of the charged sector; 
the  singlets $\Psi^n_{R,L}$ of the neutral sector dissolve into the corresponding 
three-fermion cuts,
\begin{equation}
\Psi_R\rightarrow {\cal C}[\Psi_R],\hskip0.3cm
\Psi_L\rightarrow {\cal C}[\Psi_L],\hskip0.3cm
\Psi^n_{R,L} \rightarrow {\cal C}[\Psi^n_{R,L}].
\label{scut}
\end{equation}
These three-fermion cuts ${\cal C}[\Psi_R],{\cal C}[\Psi_L]$ and 
${\cal C}[\Psi^n_{R,L}]$ are the virtual states possessing the same 
quantum numbers as three-fermion states 
$\Psi_R,\Psi_L$ and $\Psi^n_{R,L}$ so that such a dissolving process is gauge 
invariant.

Analogously, the wave-function renormalizations 
$Z_L^b(p)$ (\ref{lrenb}) and $Z_R^b(p)$ (\ref{rrenb}) of the composite scalar fields 
$\Phi_L$ and $\Phi_R$ vanish for $p\rightarrow 0$. This indicates the composite scalar
fields $\Phi_L$ and $\Phi_R$ respectively dissolve into following two-fermion cuts,
\begin{equation}
\Phi_L\rightarrow {\cal C}[\Phi_L],\hskip0.3cm\Phi_R\rightarrow {\cal C}[\Phi_R],
\label{bcut} 
\end{equation}
where ${\cal C}[\Phi_L]\sim (\psi_L, \chi_R)$ and ${\cal C}[\Phi_R]\sim (\psi_R,\chi_L)$ 
carrying the same quantum charges as $\Phi_L$ and $\Phi_R$. This is well-known as 
the composite condition of composite scalar particles\cite{wein,com}. As the composite scalar
fields dissolve into two-fermion cuts, the 1PI functions of gauge-boson couplings vanish.
In the meantime, the four-fermion coupling $a^2g_2\gg 1$ so that $\mu^2>0$ in 
eq.(\ref{bmas}) and the low-energy scaling region ${\cal A}$ is 
in the symmetric phase.

This indicates that in the low-energy scaling region located at the segment ${\cal A}$, 
all 1PI functions with external 
three-fermion states $\Psi_R,\Psi_L,\Psi^n_{R,L}$ and boson states $\Phi_L,\Phi_R$
whose energy-momenta $p$ go to zero, vanish and become irrelevant. As has been shown, 
all 1PI functions with external sterile 
neutrinos $\chi_R$ and $\chi_L$ vanish as well in the low-energy scaling region ${\cal A}$. 
As a consequence, in this low-energy scaling region, the low-energy spectra $p\sim 0$ 
are those chiral fermions $\psi_L$ and $\psi_R$ of the SM, in addition to, two free sterile 
neutrinos $\chi_R$ and $\chi_L$ . There must be an 
intermediate scale $\epsilon_c$ of the energy-momentum threshold on which 
the dissolving phenomenon occurs:
\begin{equation}
\tilde v \ll\epsilon_c \ll {\pi\over a},\hskip0.5cm \tilde v\sim 250GeV,
\label{sepsilon}
\end{equation}
where $\tilde v$ is the electroweak scale, which will be discussed in 
section (\ref{soft}). The intermediate scale $\epsilon_c$ 
is the scale of separating the massive Dirac spectra (\ref{r4l}), (\ref{r4r}), (\ref{r4ln1}) 
and (\ref{r4ln2}) from the massless chiral spectra $\psi_L$ and $\psi_R$. The 
low-energy scaling region on the segment ${\cal A}$ can be defined as,
\begin{equation}
\Omega=[0,\epsilon_c]^4,\hskip0.5cm \tilde p<\epsilon_c\ll {\pi\over 2a},\hskip0.5cm
p=a\tilde p\rightarrow 0, 
\label{continuum}
\end{equation}
in the Brillouin zone of the energy-momentum space, where $\tilde p$ is the physical 
momentum. In Eq.(\ref{sepsilon}), the value of the intermediate scale $\epsilon_c$ of the 
dissolving phenomenon depends on the value of the four-fermion coupling $a^2g_2$.

This is reminiscent of the papers\cite{wein} discussing whether helium is
an elementary or composite particle based on the vanishing of wave-function 
renormalizations of composite states. It is normally referred to as the composite condition
that the wave-function renormalizations of bound states 
go to zero ($Z\rightarrow 0$)\cite{com}. So far, we only give an intuitive and 
qualitative discussion of the dissolving phenomenon on the basis of the 
relations between the residues (generalized form factors) $Z^b_{L,R}(p)$, 
renormalized three-fermion-states
and virtual states of three individual fermions (three-fermion-cut). Evidently, we are bound to 
do some dynamical calculations to show this phenomenon could happen.

\vskip0.5cm
\noindent{\it Chiral gauge couplings.}\hskip0.3cm
Within the low-energy scaling region $\Omega$ (\ref{continuum}), all
1PI functions containing external three-fermion states vanish. This is due to the 
wave-function renormalizations $Z(p)$'s of three-fermion states vanish for
external momenta $p\rightarrow 0$. The first, 
second and last terms of gauge-fermion interacting vertices $\Lambda^L_{\mu c}(p,p')$ 
(\ref{lvdirac}) and $\Lambda^R_{\mu c}(p,p')$ (\ref{rvdirac}) vanish for $p\rightarrow 0$, 
$p'\rightarrow 0$. As a result, $\Lambda^L_{\mu c}(p,p')$ and $\Lambda^R_{\mu c}(p,p')$ 
reduce to $\Lambda^L_{\mu LL}(p,p')$ and $\Lambda^R_{\mu RR}(p,p')$. These 1PI vertex 
functions of gauge-fermion coupling vertices in the RSM 
reduce to the counterparts of the SM up to some local counterterms\cite{xue97l,xue00}.  
The Ward identities (\ref{gwardl}) and (\ref{gwardr}) are reduced to their 
counterparts of the SM,  
\begin{eqnarray}
\left({i\over a}\right)(
\sin p_\mu-\sin p'_\mu)\Lambda_{\mu LL}^L(p,p')
&=& S_{LL}^{-1}(p)-S_{LL}^{-1}(p'),\hskip0.5cm p', p\in\Omega
\label{glward}\\
\left({i\over a}\right)(
\sin p_\mu-\sin p'_\mu)\Lambda_{\mu RR}^R(p,p')
&=& S_{RR}^{-1}(p)-S_{RR}^{-1}(p'),\hskip0.5cm p', p\in\Omega
\label{grward}
\end{eqnarray}
where $S_{LL}(p)$  and $S_{RR}(p)$ are the propagators of the elementary chiral 
fermions $\psi_L$ and $\psi_R$ of the SM. These Ward identities (\ref{glward})
and (\ref{grward}) are 
consistent with the $SU_c(3)\otimes SU_L(2)\otimes U_Y(1)$ chiral gauge
symmetries realized by the chiral spectra $\psi_L$ and $\psi_R$. 

All discussions in this section are on the basis that (i) the 1PI vertex 
functions (\ref{lvdirac},\ref{rvdirac}) are computed by the strong coupling expansion 
for $p',p\simeq\pi_A$ $a^2g_2\gg 1$ in the segment ${\cal A}$ (\ref{segment}); (ii) 
the 1PI vertex functions (\ref{lvdirac},\ref{rvdirac}) are
analytically extrapolated to $p',p\sim 0$, guaranteed by the locality of the RSM. 
However, on the other hand, we directly compute 
these 1PI vertex functions in the region of $p', p\rightarrow 0$ by the weak 
coupling expansion\cite{xue99} and obtain 
consistent results. The properties discussed in this section are the crucial 
issue of the RSM, which needs to be verified by further numerical simulations.

\section{The scenario in the low-energy scaling region}\label{scenario}

We summarize the scenario (spectra and 1PI functions)
of the RSM in the low-energy scaling region $\Omega$ (\ref{continuum}) located in the segment 
${\cal A}(\ref{segment})$ in this section. Given a doublet $\psi_L$ and a singlet 
$\psi_R$ in one generation of the SM, the massive spectra in the RSM are: 

\begin{itemize} 
\item{(i)}
15 massive Dirac fermions (doublets) $\Psi_c^L$ (\ref{dlp'}) vectorially coupling to the 
gauge fields of the $SU_c^L(3)\otimes SU_L(2)\otimes U_Y^L(1)$ symmetries with the
relevant 1PI vertex function (\ref{lvdirac}) in the left-handed sector of the RSM;
 
\item{(ii)}
15 massive Dirac fermions (singlets) $\Psi_c^R$ (\ref{drp'}) vectorially coupling to 
the gauge fields of the $SU_c^L(3)\otimes U_Y^R(1)$ symmetries with the relevant
1PI vertex function (\ref{rvdirac}) in the right-handed sector of the RSM; 

\item{(iii)}
15+15 neutral massive Dirac fermions (singlets) (\ref{dnp'}) being decoupled 
consistently with the global symmetries $U_{\chi_L}(1)\otimes U_{\chi_R}(1)$ 
in the both left- and right-handed sectors of the RSM;

\item{(iv)} a massive and $SU_c^L(3)\otimes SU_L(2)\otimes U_Y^L(1)$ charged boson 
(a complex multiplet) in the left-handed sector; a massive and 
$SU_c^R(3)\otimes U_Y^R(1)$ charged boson (a complex multiplet) 
in the right-handed sector,

\end{itemize}
their mass-terms are at the lattice scale $1/a$. 

The massless fermions are:
\begin{itemize} 
\item{(i)}
the massless chiral fermions $\psi_L$ and $\psi_R$ of the SM, whose propagators are
\begin{equation} 
S^{-1}_{LL}(\tilde p)=i\gamma_\mu\tilde p^\mu\tilde Z^L_2\delta_LP_L, 
\hskip0.5cm
S^{-1}_{RR}(\tilde p)=i\gamma_\mu\tilde p^\mu\tilde Z^R_2\delta_RP_R, 
\label{sf} 
\end{equation} 
chirally coupling to gauge fields of the $SU_c(3)\otimes SU_L(2)\otimes U_Y(1)$ 
gauge group of the SM and the gauge coupling vertices 
$\Lambda^L_{\mu LL}(p,p')$ (\ref{glward}) and $\Lambda^R_{\mu RR}(p,p')$ (\ref{grward}) 
($p,p'\in \Omega$) are consistent with that of the SM up to some local counterterms;
\item{(ii)} 
two sterile chiral fermions $\chi_R\equiv\nu_R$ and $\chi_L$, whose propagators are
\begin{equation} 
S^{-1}_{nL}(\tilde p)=i\gamma_\mu\tilde p^\mu P_L, 
\hskip0.5cm
S^{-1}_{nR}(\tilde p)=i\gamma_\mu\tilde p^\mu P_R, 
\label{nsf} 
\end{equation}
being free particles.
\end{itemize}
The wave-function renormalizations $\tilde Z_2^L$ and $\tilde Z_2^R$ in eq.(\ref{sf}) 
respectively are due to the four-fermion interactions (\ref{shil}) and (\ref{shir}), 
namely, the normal mode ($p\sim 0$) of $\psi_L$ [$\psi_R$] is self-scattering via 
the weak four-fermion coupling (\ref{shil})[(\ref{shir})] without pairing up with any 
other modes\cite{xue97}. The neutral massive and massless 
fermion-sectors (\ref{dnp'},\ref{nsf}) are entirely decoupled from gauge fields.
In addition, the global chiral symmetries held in the action (\ref{stotal}) of the RSM, 
for instances, $U_{L,R}(1)$ for the fermion numbers of $\psi_L$ and $\psi_R$; 
$U_{\chi_L,\chi_R}(1)$ for the fermion numbers of $\chi_L$ and $\chi_R$ are preserved. 
However, $U_{L,R}(1)$ symmetries are 
anomalous due to the instanton effect discussed in the next section. 

High-dimension 1PI functions 
($d$=dimension) at the tree-level (in the action) can in principle induce low-dimensional 
1PI functions, If this is allowed by symmetries which are not spontaneously broken. 
The four-fermion interactions $(\ref{shil})$ and $(\ref{shir})$ given in  
the action (\ref{stotal}) are dimension-10 operators (1PI functions). We demonstrate there 
are not any 1PI functions mixing the left- and right-handed sectors of the RSM, due to the exact 
$\chi_R$- and $\chi_L$-shift-symmetries. In the both left- and right-handed sectors, we are 
interested in the relevant 1PI functions with the dimensions $d\le 4$ in the low-energy 
scaling region. The irrelevant 1PI functions with
dimensions $d>4$ vanish in the low-energy scaling region as $O(a^{d-4})$ and we are left
with the relevant 1PI functions with $d\le 4$. 

Apart from the relevant 1PI functions ($d\le 4$) with external momenta both $p\sim\pi_A$ 
and $p\sim 0$, which are intensively discussed in previous sections, 
we show as examples the 1PI vertex functions ($d>4$), induced by the four-fermion
interactions $(\ref{shil})$ and $(\ref{shir})$ in the absence of gauge fields. 
Based on the exact $SU^L_c(3)\otimes SU_L(2)\otimes U^L_Y(1)$ chiral symmetries and 
$\chi_R$-shift-symmetry in the left-handed sector of the RSM, one can straightforwardly 
obtain non-vanishing and gauge invariant 1PI functions with external momenta $p\sim \pi_A$ at
the lattice scale: the dimension-5 1PI functions are 
$\bar\psi_L\psi_L\Phi^r_L\Phi_L^{r\dagger}$, $\bar\Psi^L_c\Psi^L_c\Phi^r_L\Phi_L^{r\dagger}$ 
and $\bar\Psi^n_L\Psi^n_L\Phi^r_L\Phi_L^{r\dagger}$
($d=5$), as well as 1PI functions $d>5$.  In the right-handed sector, the counterparts of 
the non-vanishing and gauge invariant 1PI functions are obtained by replacing the subscript 
$L$ by $R$, with respect to the gauge symmetries $SU^R_c(3)\otimes U^R_Y(1)$. By adding 
the link variable $U_\mu(x)$ of 
gauge fields to the dimension-5 1PI functions, the gauge invariance of 
these 1PI functions is preserved and their dimensions are not changed. 
The gauge invariance of these 
high-dimension 1PI functions, though they are irrelevant in the low-energy scaling region,
is very important, since they can be relevant at the lattice scale. As a result,
non-perturbative (amplitude $\sim O(1/a))$ and non-smooth (correlation $\sim (O(a))$ 
variations of longitudinal gauge fields at the lattice scale are completely eliminated 
by the chiral gauge symmetric and vectorlike spectra of massive Dirac fermions.

The perturbative expansion of the link variable $U_\mu(x)$ in terms of small gauge 
fields $A_\mu$ only increases dimensions of these 1PI functions, those with $d>4$ are 
irrelevant in the low-energy scaling region $\Omega$ located at
${\cal A}$, and the SM  appears as an asymptotic chiral gauge theory of the RSM in 
the scaling limit $a\tilde p\rightarrow 0$. 

\section{Anomalies}\label{sanomaly}
  
On the basis of the scenario of fermion spectra and 1PI vertex functions presented in section
(\ref{scenario}), we discuss gauge anomalies, fermion-singlet anomalies (the $B-L$ number 
violation) and the Witten $SU_L(2)$ global anomaly. We disregard the neutral fermions 
both composite fields $\Psi^n_{L,R}$ and elementary fields $\chi_{L,R}$, because they 
completely decouple to gauge fields.

\vskip0.5cm
\noindent{\it Gauge anomalies.}\hskip0.3cm
In order to obtain the gauge anomalies produced by charged fermions, we compute the 
following $n$-point 1PI functional:
\begin{equation}
\Gamma^{(n)}_{\{\mu\}}=
{\delta^{(n)}\Gamma(A')\over\delta A'_{\mu_1}(x_1)\cdot\cdot\cdot\delta 
A'_{\mu_j}
(x_j)\cdot\cdot\cdot\delta A'_{\mu_n}(x_n)},
\label{fun}
\end{equation}
where $j=1\cdot\cdot\cdot n, (n\geq 2)$ and $\Gamma(A')$ is the vacuum
functional of the external gauge field $A'$. In terms of perturbative (amplitude $\sim O(a)$) 
and smooth (correlation $\sim O(1/a)$) gauge fields, the perturbative computations 
of the 1PI vertex functions $\Gamma^{(n)}_{\{\mu\}}$ can be straightforwardly performed by 
adopting the
method presented in ref.\cite{smit82} for the lattice QCD. Dividing the integration
of internal momenta (internal fermion loop) into 16 hypercubes in the Brillouin zone.
The first hypercube where the chiral fermions $\psi_L$ and $\psi_R$ live is the low-energy 
scaling region $\Omega$ 
given by eq.(\ref{continuum}), while the other 15 hypercubes,  at edges of the Brillouin zone,
where massive Dirac fermions $\Psi_c^L$ (\ref{dlp'}) and $\Psi_c^R$ (\ref{drp'}) live
are given by ${\cal D}'s$,
\begin{equation}
{\cal D}'s:\hskip0.5cm p\sim \pi_A.
\label{dregion}
\end{equation}
We totally have 16 contributions to the truncated n-point 1PI functional (\ref{fun}).  

Analogous to the massive Dirac doublers in the lattice QCD\cite{smit82}, the massive 
spectra of charged 
Dirac fermions $\Psi_c^L$ (\ref{r4l},\ref{dlp'}) and $\Psi_c^R$ (\ref{r4r},\ref{drp'}) 
in ${\cal D}'s$  are vectorlike consistently 
with the gauge symmetries $SU_c^L(3)\otimes SU_L(2)\otimes U_Y^L(1)$ and 
$SU_c^L(3)\otimes U_Y^R(1)$ respectively in the left- and right-handed sectors of 
the RSM. By using a continuous 
regularization in each hypercube of ${\cal D}'s$, we make perturbative computations 
(\ref{fun}), analogously to the ananlysis in ref.\cite{xue00}, and show these 
vectorlike and massive Dirac fermions do not have any contributions 
to non-local gauge anomalies. This is a nature result from the vectorlike Ward
identities (\ref{gwardl}) and (\ref{gwardr}). Beside, these massive Dirac fermions 
and their relevant 1PI functions preserve both $U_L(1)$ and $U_R(1)$ symmetries, 
thus do not contribute to the fermion-singlet anomalies. However, computations give rise to
some finite local gauge variant terms that are due to the 
divisions of the Brillouin zone into hypercubes.  
  
In the first hypercube $\Omega$ (\ref{continuum}), to obtain the gauge anomalies attributed 
to the chiral fermions $\psi_L$ and $\psi_R$, we perturbatively compute 1PI-functional
(\ref{fun}) by adopting the Pauli-Villars
regularization with finite number of Pauli-Villars massive 
fermion regulators at the scale $\epsilon_c$.  
The non-trivial contributions to the gauge anomalies are given by
\begin{eqnarray}
\Gamma^{i=1}_{\mu\nu\alpha}(p,q)\!&\!=\!&\!\int_\Omega\!
{d^4k\over(2\pi)^4}\tr\!\left[S(k\!+\!{p\over2})\Gamma_\mu(k)S(k\!-\!{p\over2})
\Gamma_\nu(k\!-\!{p\!+\!q\over2})S(k\!-\!{p\over2}\!-\!q)\Gamma_\alpha(k\!-\!
{q\over2})\right]\nonumber\\
&&+(\nu\leftrightarrow\alpha),
\label{tri}
\end{eqnarray}
where the propagators $S(p)=S_{LL}(p),S_{RR}(p)$ and vertices 
$\Gamma_\mu=\Lambda_{\mu LL}^L,\Lambda_{\mu RR}^R$ given by eqs.(\ref{glward},\ref{grward}).
As a result, modulo 
possible finite local counterterms, we obtain the consistent gauge anomalies 
for the $SU_L(2)\otimes U_Y(1)$ symmetries, which are the same as the continuum counterpart and
proportional to,
\begin{equation}
\sum_{\rm doublets}{\rm tr}(T^aT^aY_L)=0,
\label{anomaly}
\end{equation}
where $T^a$ is a component of the $SU_L(2)$ weak isospin and $Y_L$ denotes the 
$U_Y(1)$ weak hypercharge of the doublets $\psi_L$ in the left-handed sector, comprising 
the left-handed lepton and quark doublets. This gauge anomalies (\ref{anomaly}) vanish due 
to the fermion content of the SM. The gauge anomalies of $SU^{L,R}_c(3)$ gauge 
symmetries of quark fields are canceled between the left- and right-handed quark 
sectors. Other contributions containing anomalous vertices
$(\psi\bar\psi AA, \psi\bar\psi AAA)$ vanish within 
the hypercube $\Omega =[-\epsilon_c,\epsilon_c]^4$. 

As already mentioned, the finite local gauge variant terms are raised from 
perturbative calculations by using the Pauli Villars regularization in each 
hypercube of the Brillouin zone.  These finite local gauge variant terms called as 
the residual breakings $R(\epsilon_c)$ are 
gauge symmetry-breakings at the scale of $O(\epsilon_c)$, rather than the lattice 
scale $O(\pi/ a)$. These residual breakings $R(\epsilon_c)$ are only relevant to 
the external gauge fields (the transverse and longitudinal 
components) with small fluctuations ($O(\epsilon_c)$) and smooth correlations 
($O(1/\epsilon_c)$) at the scale $\epsilon_c\ll \pi/a$ (\ref{sepsilon}). 
Since the gauge anomalies are canceled within the fermion content and the 
vacuum functional (\ref{fun}) 
is determined up to finite local counterterms, we are allowed to
self-consistently add appropriate finite local counterterms to eliminate these 
residual breakings $R(\epsilon_c)$, in order to 
achieve an asymptotically chiral gauge field theory for the SM in the low-energy limit.  
This is the same as the procedures in the normal renormalization prescription 
of quantum field theories in the continuum. We would like to point out that
these residual breakings $R(\epsilon_c)$ produced artifically by using the Pauli-Villars 
regularizations in each hypercube of the Brillouin zone must exactly
cancel each other on the basis of 
the arguments that (i) gauge symmetries are exactly preserved; (ii) all 1PI functions are
analytically continuous from one hypercube to another for the locality of the RSM. 
However, this has to be rigorously proved.

To be consistent with the manifest chiral gauge symmetries of the RSM
both the action (\ref{stotal}) and fermionic measure,
the gauge anomaly (\ref{anomaly}) and residual breakings $R(\epsilon_c)$ must 
be canceled within the fermion content of the 
theory. Otherwise, the vectorlike spectrum of fermion zero modes must appear, either 
doublers do not decouple or the three-fermion-states $\Psi_R$ and $\Psi_L$ do not 
dissolve into their three-fermion-cuts and become massless Weyl particles in the 
low-energy region. From this point of view, we see the anomaly-cancelation by the 
fermion content is a necessary condition for this scenario to work.

\vskip0.5cm
\noindent{\it Fermion number violations.}\hskip0.3cm
To compute the $B+L$ number violation, i.e., the flavour-singlet anomaly due to the $SU_L(2)$ 
instanton effect, we can adopt the approach of computing the mixing 
anomalies\cite{xue00,preskill91}. Because the sterile neutrinos $\chi_L$ and $\chi_R$ are
completely decoupled from gauge fields (see eqs.(\ref{nlcc}) and (\ref{nrcc})), the global 
symmetries $U_{\chi_L}(1)$ and $U_{\chi_R}(1)$
cannot be mixed with the $SU_L(2)$ as a commuting $U(1)$ factor in the $SU_L(2)$ gauge group. 
Therefore, there is no any mixing anomalies between $U_{\chi_L,\chi_R}(1)$ and 
$SU_L(2)$, otherwise the sterile neutrinos $\chi_L$ and $\chi_R$ would not be 
decoupled from gauge fields (cf. section 6 in ref.\cite{xue00}).

Let us first consider the lepton sector: doublets $\psi_L^l$ and singlets $\psi_R^l$. The 
$U(1)$ symmetry for the lepton number $U^l(1)=U_L(1)=U_R(1)$ in the lepton sector. 
$U^l(1)$ associates 
to the flavour-singlet Noether current of leptons:
\begin{equation}
j^L_\mu = i\bar\psi_L^l\gamma_\mu\psi^l_L+i\bar\psi_R^l\gamma_\mu\psi^l_R;\hskip0.5cm
\partial^\mu j^L_\mu(x)=0,
\label{lcu}
\end{equation}
where the conservation of the current $j^L_\mu(x)$ is held at the tree level, since the action 
(\ref{stotal}) of the RSM processes $U_L(1)$ and $U_R(1)$ global symmetries. 
Eq.(\ref{lcu}) corresponds to the conservation of the lepton numbers. However, as we know, 
eq.(\ref{lcu}) should be anomalous. We compute this anomaly by using
the ``mixing'' gauge group $SU_L(2)\otimes U^l(1)$. Following computations (\ref{tri}) 
for the gauge anomaly (\ref{anomaly}), we just need to replace the hypercharge $Y$ in 
(\ref{anomaly}) by the generator of $U^l(1)$ for the lepton sector. In the first hypercube $\Omega$, actually, only 
the left-handed doublets of leptons have contributions to the anomaly. No any 
contributions are stemming from the hypercubes ${\cal D}$ (\ref{dregion}) where Dirac 
fermions are vectorlike. Up to gauge invariant local counterterms, we obtain 
\begin{equation}
\partial^\sigma j^L_\sigma = {i\over32\pi^2}\tilde F^{\mu\nu}F_{\mu\nu},
\label{la}
\end{equation}
where $F$ is the field strength of the $SU_L(2)$ gauge group.

Then we consider the quark sector: doublets $\psi_L^q$ and singlets $\psi_R^q$. The 
$U(1)$ symmetry for the baryon number $U^b(1)=(U_L(1))^3=(U_R(1))^3$ for the 
quark sector. $U^b(1)$ associates 
to the flavour-singlet Noether current of quarks:
\begin{equation}
j^B_\mu = i\sum_q\left(\bar\psi_L^q\gamma_\mu\psi^q_L+\bar\psi_R^q\gamma_\mu\psi^q_R\right);
\hskip0.5cm \partial_\mu j^B_\mu(x)=0,
\label{qcu}
\end{equation}
where $\sum_q$ is the summation of all colours and the conservation of the current $j^q_\mu(x)$ 
is held at the tree level, since the action 
(\ref{stotal}) of the RSM processes $U_L(1)$ and $U_R(1)$ global symmetries. Eq.(\ref{qcu}) 
corresponds to the conservation of the baryon numbers, which is 3 times of the quark 
numbers. However, as we know, eq.(\ref{qcu}) should also be anomalous. We compute this 
anomaly by using the ``mixing'' gauge group $SU_L(2)\otimes U^q(1)$. Analogously, in the 
computations (\ref{tri}) for the gauge anomaly (\ref{anomaly}) and we replace the hypercharge 
$Y$ in (\ref{anomaly}) by the generator of $U^b(1)$, and sum over colours in
the quark sector. Analogously, up to gauge invariant 
local counterterms,  we obtain 
\begin{equation}
\partial^\sigma j^B_\sigma = {i\over32\pi^2}\tilde F^{\mu\nu}F_{\mu\nu}.
\label{qa}
\end{equation}
These results (\ref{la}) and (\ref{qa}) lead to the $B-L$ number conservation 
and $B+L$ number violation,
\begin{eqnarray}
\partial^\sigma (j^B_\sigma-j^L_\sigma)&=&0\nonumber\\ 
\partial^\sigma (j^B_\sigma+j^L_\sigma)&=&{i\over16\pi^2}\tilde F^{\mu\nu}F_{\mu\nu}.
\label{b-l}
\end{eqnarray}

We emphasize crucial points for achieving the correct form of the $B+L$ number violation
(\ref{b-l}) up to gauge invariant local counterterms in this scenario. (i) The 
composite three-fermion-states $\Psi_R$ and $\Psi_L$ dissolve into their 
three-fermion-cuts ${\cal C }[\Psi_R]$ and ${\cal C }[\Psi_L]$, in another word, no massless
composite three-fermion-states $\Psi_R$ and $\Psi_L$ exist in the low-energy 
scaling region $\Omega$ (\ref{continuum}). (ii) The
complete decoupling of the massless sterile fermions $\chi_R$ and $\chi_L$ from 
gauge fields, which leads to the current conservations (\ref{nlcc},\ref{nrcc}). 
(iii) The exact chiral gauge symmetry for both the gauge anomaly (\ref{anomaly}) 
and residual breakings $R(\epsilon_c)$ are eliminated.  

In the fermion content of the SM, we certainly have the possibilities of fermion-number
violating, chiral gauge symmetric four-fermion couplings at the lattice scale. The nice 
examples are given in refs.\cite{ep,mc}. The low-energy scaling region for the desired 
gauge symmetric spectrum of the SM could be achieved and 
these fermion-number violating four-fermion couplings would be relevant 
operators in such a scaling region to give the $B+L$ fermion number violation.

Before ending this section, we mention that the $SU_L(2)$ global anomaly of the Witten 
type\cite{witten} appears for odd numbers of left-handed $SU_L(2)$ doublets. Thus, we 
do not find any inconsistency due to the Witten anomaly in the RSM, 
since there are even numbers of left-handed $SU_L(2)$ doublets $\psi_L$ for 
leptons and quarks in the low-energy scaling region.

\section{Soft breakings}\label{soft}

\vskip0.5cm
\noindent{\it Residual breakings.}\hskip0.3cm
In the RSM, both the action (\ref{stotal}) at the lattice scale and the low-energy 
scaling region ${\cal A}$ (\ref{segment}) are exactly chiral-gauge invariant.
The non-perturbative fluctuations of longitudinal gauge fields at the lattice scale
are thus eliminated. 
Nevertheless, in the practice of perturbatively computing 1PI functions in terms of small 
gauge fields, we have to artificially divide the whole Brillouin zone into 16 hybercubes
and introduce a continuous regularization scheme 
which breaks chiral-gauge symmetries at the intermediate scale $\epsilon_c$. As a 
result, residual breakings $R(\epsilon_c)$ are produced. These residual breakings 
$R(\epsilon_c)$ are canceled by adding appropriate counterterms. This is reminiscence 
of the two-cutoff approach to chiral gauge theories on the lattice\cite{thooft}. 
Though such a practical framework certainly 
is self-consistent, we speculate that the RSM should not intrinsically 
have any residual breakings at any scale for its exact gauge invariance and locality. 
This drives our attention to recent approaches \cite{nerev,lurev} to chiral gauge theories,
which have been greatly developed in recent literatures\cite{recent}.   

In these approaches, the regularizations of the chiral fermion sector use 
the Ginsparg-Wilson equation\cite{gw} that was obtained from the renormalization 
group equation approaching to a low-energy scaling region. Owing to the Dirac operator 
satisfying the Ginsparg-Wilson equation, the residual breakings of the gauge symmetry 
are reduced\cite{gs99}. This implies that the residual breakings originated from 
the Ginsparg-Wilson Dirac operator are high-dimension (non exactly local) operators, 
which are becoming to be irrelevant in the renormalization flow approaching to the scaling 
region. Furthermore, with the Ginsparg-Wilson Dirac operator, it is proved \cite{lurev} 
that all residual breakings of 
the gauge symmetry and the gauge anomaly can be rewritten as a total divergence 
for its topological nature, thus they are eliminated by redefining the gauge current
\cite{lurev} for the finite lattice spacing and without fine-tuning. This 
implies that the renormalization scaling flow governed by the renormalization group 
equation is gauge invariant, as a consequence, demonstrating the existence of a gauge 
invariant low-energy scaling region for chiral-gauge invariant spectra and 1PI functions. 
 
In the RSM action (\ref{stotal}), the four-fermion couplings are made to be exactly local 
and chiral gauge symmetric at the lattice scale. We can formally integrate away 
the sterile neutrino fields $\chi_R, \chi_L$ and obtain the effective Dirac actions  
bilinearly in the fermion fields $\psi_L$ and $\psi_R$. Such effective Dirac actions 
are obviously not exactly local. It is worthwhile to examine whether such effective 
Dirac operators could be in the same universal class of the solutions to the Ginsparg-Wilson 
equation in the sense of the renormalization group invariance. It is also important 
to examine that in the low-energy scaling region ${\cal A}$ (\ref{segment}),
the relevant spectra and 1PI functions induced from the four-fermion couplings of the RSM 
are in the same universal class of the relevant spectra and 1PI functions originated 
from the Ginsparg-Wilson Dirac operators in the view of the renormalization group invariance.
For simplicity, this examination should be first carried out in a simple two-dimensianl model, 
such as the (11112)-model\cite{11112}. 
We expect that in our approach (RSM) and the approach based on the Ginsparg-Wilson 
equation, the low-energy relevant spectra and 1PI functions should be the same, the difference
should be irrelevant high-dimension operators in high-energy range and these irrelevant 
high-dimension operators should be vanishing 
($O(a^n)n>1$), as the renormalization flow approaches the low-energy scaling region ${\cal A}$.
If this expectation can be demostrated, two important and perspective points are followed:
(i) many analytical developments in the approached based on the Ginsparg-Wilson equation\cite{nerev,lurev,recent} should also be discussed in the framework of our approach;      
(ii) the effective Dirac action after integrating over the sterile neutrino fields in our approach
can be adopted for non-perturbative numerical simulations, and the low-energy scaling region 
${\cal A}$ can be adopted as the scaling region for numerical simulations to approach the continuum limit.    

In fact, on the basis of our studies and recent approaches based on the 
Ginsparg-Wilson equation, it can be strongly concluded that we can not only
find a gauge invariant regularization for chiral gauge theories (like the SM) at 
the lattice scale, but also a gauge invariant low-energy scaling region for the 
desired low-energy spectra and 1PI functions of chiral gauge theories. This
has been generally believed to be impossible and we nevertheless
have persisted in working for a decade. The studies of an appropriate universal 
class of gauge invariant four-fermion 
couplings at the lattice scale, which processes the desired low-energy scaling region 
for the SM, are highly deserved and far-reaching, since they could be 
Nature's choice for the extensions of the SM to the high-energy region. 

\vskip0.5cm
\noindent{\it Electroweak scale.}\hskip0.3cm
As mentioned in the introductory section, the Higgs sector of the SM is disregarded and 
the main goal of this paper is to study the RSM: a gauge invariant 
regularization of the SM. RSM can possibly achieve a gauge-invariant low-energy 
scaling region 
where the chiral fermion spectra and gauge couplings of the SM are realized.
As has been proved, the $\chi_R$- and $\chi_L$-shift symmetries of the RSM rigorously 
protect the decoupling between the left- and right-handed sectors of the RSM. 
Both hard and soft spontaneous symmetry-breakings are strictly prohibited. 
Thus all fermions are exactly massless. However the soft spontaneous symmetry-breaking 
is needed for the fermion mass generation. It is conceivable that such a soft spontaneous 
symmetry-breaking should 
be allowed, provided the scale of the soft spontaneous symmetry-breaking is well 
below the intermediate dissolving scale $\epsilon_c$ (\ref{sepsilon}).  

The electroweak scale $\tilde v$ is the soft spontaneous symmetry-breaking in the RSM.
It is related to the non-zero v.e.v. 
$\langle\bar\psi_L\psi_R\rangle$, which can be induced by the dimension-6 four-fermion 
interactions of the Nambu Jona-Lasinio type\cite{njl},
\begin{equation}
G\bar\psi_L\cdot\psi_R\bar\psi_R\cdot\psi_L,
\label{njl}
\end{equation} 
where the NJL four-fermion coupling $G$ has to be at the intermediate 
scale $\epsilon_c$, rather than the lattice scale. The NJL four-fermion interactions 
(\ref{njl}) mix left- and right-handed fermions of the RSM. Obviously, the $\chi_R$- and 
$\chi_L$-shift symmetries of the RSM are explicitly broken.  It is required that the NJL 
interactions (\ref{njl}) induces 1PI functions, for example the dimension-3 mass 
operator $\bar\psi_L\cdot\psi_R$, that are only relevant up to the electroweak scale $\tilde v$. 
Otherwise, the $\chi_R$- and 
$\chi_L$-shift symmetries of the RSM would be strongly violated and the 
scenario of the RSM in the low-energy scaling region is jeopardized.
We speculate that this could be achieved by finding a ultra-violet stable point 
of the scaling value $G=G_c$ in the low-energy scaling region ${\cal A}$ 
(\ref{segment}) so that (i) the soft 
spontaneous symmetry-breaking $\tilde v$ is much small than $\epsilon_c$ and dimension-3 mass 
functions $\bar\psi_L\cdot\psi_R$ are irrelevant at the scale $\epsilon_c$; (ii) the 
four-fermion interactions (\ref{njl}) become effectively relevant and 
renormalized dimension-4 operators by receiving non-perturbative wave-function 
renormalization. If this could be realized, the scenario of the RSM would be not only 
remained, but also enriched with low-lying massive fermions.       
We leave the further discussions and computations in a separate paper. 

\section{Restoration of the parity-symmetry}\label{restore}

As discussed in section (\ref{scenario}), in the scaling region $\cal A$ 
(\ref{segment}), the RSM gives rise to not only the desired 
chiral-fermion spectra and 1PI functions of the SM in the low-energy region, but also the 
massive Dirac fermions and vectorially gauge-symmetric 1PI functions in the high-energy region, 
characterized by the energy-momentum threshold $\epsilon_c$ (\ref{sepsilon}). 
The most striking feature 
of the RSM is that the parity-violating gauge symmetries in the low-energy region turn to the 
parity-conservation gauge symmetries in the high-energy region above the 
energy-momentum threshold $\epsilon_c$, 
without any other extra elementary fermions and gauge bosons. However, {\it a priori}, 
the RSM cannot determine the definite value of the threshold $\epsilon_c$, which must 
be much larger than the electroweak scale, as indicated by current experimental 
results. This restoration of the parity-conservation gauge symmetries could 
be experimentally tested by measuring the left-right asymmetry,
\begin{equation}
A_{LR}={\sigma_L-\sigma_R\over\sigma_L+\sigma_R}
\label{lras}
\end{equation}
at the high-energy $\ge\epsilon_c$ above the electroweak scale, where $\sigma_L$ and 
$\sigma_R$ respectively denote the cross sections for an incident left-handed 
and right-handed polarized electrons. The left-right 
asymmetry $A_{LR}$ is related to the 1PI vertex functions of gauge couplings to fermions.
The non-vanishing $A_{LR}$ indicates parity-violating gauge-fermion interactions, while 
vanishing $A_{LR}$ indicates parity-conservation gauge-fermion interactions. The scenario
of the RSM predicates that in the high-energy region the left-right asymmetry $A_{LR}$ 
should be vanishing and no extra very massive gauge bosons can be detected, 
differently from the left-right symmetric models\cite{lrmodel}.

In addition to the restoration of parity-conservation symmetries, the vectorially 
gauge-symmetric 1PI functions of gauge-fermion interactions indicate that 
the $W^\pm$-bosons have not only gauge coupling to left-handed fermions, but also 
gauge coupling to right-handed fermions in the high-energy region. This clearly 
modifies the low-energy physics of the SM and possibly gives some clues to the flavour physics. 
Actually, with the vectorlike gauge-coupling vertex of the $W^\pm$-bosons, we had done some 
preliminary studies of phenomenological aspects on the generation of fermion masses, 
fine-tuning problem, masses and 
mixing angles of the quark and lepton sectors\cite{neu00}. We speculate that
in the neutral sector of the RSM, the light and sterile neutrinos $\chi_R$ and $\chi_L$ can 
be candidates of the part of the hot dark matter, while massive and neutral Dirac fermions 
$\Psi_n^L$ and $\Psi^n_R$ could be candidates for the part of the cold dark
matter; in the charged sector, the massive charged Dirac fermions $\Psi_c^L$ and $\Psi_c^R$ 
could decay to produce high-energy cosmic rays. The right-handed neutrino $\chi_R\equiv\nu_R$ 
and sterile left-handed neutrinos $\chi_L$ should play their roles on the neutrino physics. 
    
\section{Summary and conclusions}

The phenomenon of vectorlike fermion 
doubling described by the ``no-go'' theorem of Nielsen and Ninomiya is very
generic, provided the degree of freedoms of a gauge-invariantly regularized field 
theory is finite. The paradox is thus raised: vectorially gauged fermions and their doublers 
must appear in a naive gauge-invariantly regularized SM, on the contrary, chiral gauged 
fermions are in fact the fermion content of the SM. By introducing two sterile neutrinos, 
we propose four-fermion interactions and dynamics leading to a possible resolution of this 
paradox.  We discuss the non-perturbative 
regularizations for the SM without the Higgs sector, i.e., the RSM (\ref{stotal}). The analysis 
and discussions of fermion spectra, 1PI functions and anomalies in the gauge invariant 
low-energy scaling region (\ref{segment}) are given, analogously to that presented in 
refs.\cite{xue97,xue99,xue00} for the $SU_L(2)$ chiral-gauged fermions on the lattice. We 
conclude that the RSM and its dynamical scenario provide a plausible resolution to 
the paradox: the 
long-standing problem of gauge-invariantly regularizing chiral gauge theories, 
like the SM on a lattice. We find  
a universal class of effective four-fermion interactions, whose gauge invariant scaling 
region for the SM appearing as an asymptotic chiral gauge theory in the low-energy region, 
while parity-conservation gauge symmetries are restored in the high-energy region. 

On the other hand, we have to confess that it is very necessary and inviting
for numerical computations and other techniques to verify the RSM and its dynamical scenario, 
in particular, the phenomenon of no hard spontaneous symmetry-breaking at the lattice scale and 
the phenomenon of the three-fermion-states gauge-invariantly dissolving into 
corresponding three-fermion-cuts at the intermediate energy-threshold 
$\epsilon_c$ (\ref{sepsilon}). 

The ultimate regularization of the fundamental quantum field 
theory for particle physics must be given by the nature regulator -- the quantum gravity
with the fundamental Planck scale, it is interesting and important to search for the 
effective high-dimension operators for regularizing the $SU(5)$ and $SO(10)$ 
chiral gauge theories that unify three types of the gauge-fermion interactions of the SM at 
the GUT scale.  

To end this article, we wish to make a very general and brief discussion on any possible relationships between the multifermion coupling and bilinear fermion coupling approaches for anomaly-free chiral gauge theories on the lattice. 
In both bilinear fermion and multifermion coupling models, extra fermionic species must be decoupled and right-handed and left-handed fermionic species must couple in some ways to have anomalies. The couplings between right- and left-handed fermionic species appear either in the action or in the fermionic measure. 

The Wilson fermion is exact local (in the range of the lattice spacing) and  right- and left-handed fermionic species couple at the lattice scale. Doublers are very massive and decoupled. As required by the ``no-go" theorem, the residual breakings of the gauge symmetry are at the lattice scale. These residual breakings of the chiral gauge symmetry can be eliminated by adding and fine-tuning appropriate counterterms so as to enforce the Ward identities associated to exact chiral gauge symmetries at the continuum limit\cite{rome}. 

Contrasting with the Wilson fermion, the regularization of the fermion sector adopted by the ``overlap"\cite{nerev} and L\"uscher\cite{lurev} approaches use the Ginsparg-Wilson equation\cite{gw} that was obtained from the renormalization group equation. Owing to the Dirac operator satisfying the Ginsparg-Wilson equation, the residual breakings of the gauge symmetry are reduced\cite{gs99} and supposed to be eliminated by either average over gauge configurations or adding local counterterms in order to preserve exact chiral gauge symmetries. In the L\"uscher\cite{lurev} approach for the abelian gauge theory, all residual breakings of the gauge symmetry including the gauge anomaly can be rewritten as a total divergence for its topological nature, thus they are eliminated by redefining the gauge current\footnote{Private conmmunication with L\"uscher.} for the finite lattice spacing and without fine-tuning. In order to completely decouple extra fermion species, as required by the ``no-go" theorem, the approaches relax the exact locality to the locality whose range extends  to a few lattice spacings with an exponential tail. 

While in the models of multifermion couplings, the couplings of right-handed and left-handed fermions can be made exactly local and chiral gauge symmetric at the lattice scale. However, as discussed at the end of section 2, we have to find a peculiar multifermion coupling and a scaling region desired for the low energy. The hard spontaneous symmetry breaking is absolutely not tolerated so that residual breakings of gauge symmetries are not at the lattice scale. The strong coupling at the high energy is needed and three-fermion cut must be realized at the low-energy so as to decouple extra fermion species with ``wrong" chirality. 

Taking our action (\ref{stotal}) as an example, we can formally integrate away the spectator field $\chi_R$ and obtain the effective Dirac action  bilinearly in the fermion field $\psi_L^i$. Such an effective Dirac action is obviously not exactly local. It is worthwhile to examine whether such an effective Dirac operator could be the solution to the Ginsparg-Wilson equation in the sense of the renormalization group invariance. Most importantly, we need to show in the scaling region ${\cal A}$ (\ref{segment}) for the low-energy, whether the relevant spectra and operators induced from the multifermion couplings(high dimension operators) at the lattice scale are in the same universal class with the solutions to the Ginsparg-Wilson equation, in the view of the renormalization group invariance. 

In fact, the recent successful progress based on the Ginsparg-Wilson equation strongly implies the existence of the scaling region \ref{segment}) for exactly chiral-gauge symmetric theories in the low-energy, obtained from our model (\ref{stotal}). This has been generally believed to be impossible. 
The studies of appropriate multifermion couplings at the lattice scale and desired scaling region are highly deserved, since they could be Nature's choice for chiral gauge theories, e.g., the Standard Model, at the high-energy.



\begin{references}

\bibitem{nn81}
H.B.~Nielsen and M.~Ninomiya, Nucl.~Phys.~B185 (1981) 20, {\it
ibid.} {\bf B193} (1981) 173.

\bibitem{nn81plb}
H.B.~Nielsen and M.~Ninomiya, Phys.~Lett. B105 (1981) 219.

\bibitem{nn91}
H.B.~Nielsen and M.~Ninomiya, Int.~J.~ of Mod.~Phys. A6 (1991) 2913.

\bibitem{gw}
P.H.~Ginsparg and K.G.~Wilson, Phys. Lett. D25 (1982) 2649.

\bibitem{ep}
E.~Eichten and J.~Preskill, Nucl.~ Phys. B268 (1986) 179.

\bibitem{ss}
J.~Smit, Acta Physica Polonica B17 (1986) 531;\\
P.D.V.~Swift, Phys.~Lett. B145 (1984) 256.

\bibitem{aoki}
S.~Aoki, Phys.~Rev.~Lett. 60 (1988) 2109;
Phys.~Rev. D38 (1988) 618; Nucl.~Phys. (Proc.~Suppl.) 
B29 (1992) 71;\\
T.D.~Kieu, D.~Sen and S.-S.~Xue, Phys.~Rev.~Lett. 61 (1988) 282.

\bibitem{ahx}
S.~Aoki, I-Hsiu Lee and S.-S.~Xue, Phys.~Lett. B229 (1989) 79
and BNL Report 42494 (1989).

\bibitem{gp}
M.F.L.~Golterman, D.N.~Petcher,  Phys.~Lett. B225 
(1989) 159.

\bibitem{px91}
G.~Preparata and S.-S.~Xue, Phys.~Lett. B264 (1991) 35.

\bibitem{rome} 
A.~Borrelli, L.~Maiani, G.C.~Rossi, R.~Sisto and M. Testa, Nucl.~ Phys.
B333 (1990) 335; Phys.~Lett. B221 (1989) 360;\\
L.~Maiani, G.C.~Rossi, and M. Testa, Phys.~Lett. B261 (1991) 479;
{\it ibid} B292 (1992) 397;\\
L.~Maiani, Nucl.~ Phys. (Proc.Suppl.) 
B29 (1992) 33.\\
J.L. Alonso, Ph. Boucaud, F. Lesmes and A.J. van der Sijs
 Nucl.~ Phys.
B457 (1995) 175 and {\it ibid} 472(1996)738 (ERRATUM).

\bibitem{terev}
M. Testa, The Rome approach to chirality,
Talk given at the APCTP - ICTP Joint International Conference (AIJIC 97) 
on Recent Developments in
Nonperturbative Quantum Field Theory, Seoul 1997,
hep-lat/9707007

\bibitem{monrev}
I.~Montvay, Phys.~Lett. B199
(1987) 282; Nucl. Phys. B29 (Proc.~Suppl.)
(1992) 159, {\it ibid} 30B (1993) 621 
and references therein.

\bibitem{perev}
D.N.~Petcher, Nucl.~Phys. (Proc.~Suppl.) B30 
(1993) 52, references there in.

\bibitem{pgr}
M.F.L.~Golterman, D.N.~Petcher and E.~Rivas, Nucl.\ Phys. B395 
(1993) 597.

\bibitem{shrev}
Y.~Shamir, Phys.~Rev.~Lett. B71 (1993) 2691; hep-lat/9307002
Nucl.~Phys. (Proc.~Suppl.) B34 (1994) 590, {\it ibid} B47 (1996) 212,
references there in.

\bibitem{px}
G.~Preparata and S.-S.~Xue, Phys.~Lett. B335 (1994) 192, {\it ibid} B329 (1994) 87,
B325 (1994) 161, B377 (1996) 124,  B395 (1997) 257; Nucl.~Phys. B26 (Proc.~Suppl.) 
(1992) 501; {\it ibid} B30 (1993) 647.\\
S.-S.~Xue, Phys.~Lett. B313 (1993) 411, {\it ibid}
B335 (1994) 192, B377 (1996) 124; Nucl.~Phys.(Proc.~Suppl.) B47 (1996) 583.

\bibitem{ksymir} 
D.~Kaplan, Phys.~Lett. B288 (1992) 342;\\
M.F.L.~Golterman, K.~Jansen and D.~Kaplan Phys.~Lett. B301
(1993) 219;\\
M.F.L.~Golterman, K.~Jansen, D.N.~Petcher, and J.C.~Vink, Phys.~Rev.
D49 (1994) 1606;\\
M.F.L.~Golterman and Y.~Shamir Phys.~Rev. D51 (1995) 3026;\\
Y.~Shamir, Nucl.~Phys. B406
(1993) 90, {\it ibid} B417 (1994) 167.

\bibitem{thooft}
M.~Gockeler, A.~Kronfeld, G.~Schierholz and U.~Wiese,
Nucl.~ Phys. B404 (1993) 839;\\
G.~'t Hooft, Phys.~Lett. B349 (1995) 491;\\
P.~Hern\'andez and R.~Sundrum, Nucl.~ Phys. B455 (1995) 287;\\
G.~Bodwin, Phys.~Rev. D54 (1996) 6497.

\bibitem{slanov} 
S.~Frolov and A.~Slavnov, Nucl.~ Phys. B411 (1994) 647.

\bibitem{crrev}
M.~Creutz, Nucl.~Phys. (Proc.~Suppl.) B42 (1995) 56,
Phys.~Rev. D52 (1995) 2951, hep-lat/9912006, talk at the conference "Lattice 
Fermions and structure of the vacuum" October 5-9, 1999, Dubna, Russia;\\
M.~Creutz and I.~Horv\'ath,
Phys.~Rev. D50 (1994) 2297.

\bibitem{over}
R.~Narayanan and H.~Neuberger, Nucl.~ Phys. B443 (1995) 305, 
and references there in.

\bibitem{xue96}
S.-S.~Xue, Phys.~Lett. B381 (1996) 277.

\bibitem{xue97}
S.-S.~Xue, Nucl.~Phys. B486 (1997) 282.

\bibitem{xue97l}
S.-S.~Xue, Phys.~Lett. B408 (1997) 299; 
Nucl.~Phys. (Proc.~Suppl.) B53 (1997) 668.

\bibitem{mc}
M.~Creutz, C.~Rebbi, M.~Tytgat and S.-S.~Xue,
Phys.~Lett. B402 (1997) 341;\\
M.~Creutz,  Nucl.Phys.Proc.Suppl. 63 (1998) 599;\\
S.-S.~Xue, Nucl.Phys.Proc.Suppl. 63 (1998) 596.

\bibitem{nerev}
H.~Neuberger, Nucl.~Phys.~Proc.~Suppl. 83 (2000) 67-76, Chin.~J.~Phys. 38 (2000) 735-743, 
{\it ibid} 38 (2000) 533-542, hep-lat/0011035, Ann.Rev.Nucl.Part.Sci. 51 (2001) 23-52, 
Phys.Rev. D63 (2001) 014503, hep-lat/0303009.

\bibitem{lurev}
Martin L\"uscher, Phys.Lett. B428 (1998) 342-345; Nucl.~Phys.~Proc.~Suppl. 83 (2000) 34-43,
Nucl.Phys. B538 (1999) 515-529,{\it ibid} B549 (1999) 295-334, B568 (2000), {\it ibid} 162-179, 
JHEP 0006 (2000) 028,
hep-th/0102028, Lectures given at the International School of Subnuclear Physics, Erice, 27 August - 5 September 2000, and references there in.

\bibitem{gs99}
Wolfgang Bock, Maarten Golterman, Ka Chun Leung, Yigal Shamir, hep-lat/9912025,
contribution to workshop ``Lattice Fermions and Structure of the Vacuu'' 
(Dubna, Russia), eds.~ V.~Mitrjushkin, G.~Schierholz (Kluwer, 2000), 
and references there in.

\bibitem{testa99}
M. Testa,  hep-lat/9912035, Chin.~J.~Phys. 38 (2000) 563-572,
and references there in.

\bibitem{recent}
W.~Kerler,  Int.J.Mod.Phys. A16 (2001) 3117-3150,  Phys.Lett. B510 (2001) 325-328, 
JHEP 0210 (2002) 019, Nucl.Phys. B646 (2002) 201-219, Int.J.Mod.Phys. A18 (2003) 2565-2590;\\
T.~W.~Chiu, Chin.J.Phys. 38 (2000) 573-582, Nucl.Phys. B588 (2000) 400-416, Phys.Lett. B521 (2001) 429-433, Phys.Rev. D65 (2002) 054508;\\
K.~ Fujikawa, Chin.J.Phys. 38 (2000) 551-562, Nucl.Phys. B589 (2000) 487-503, hep-lat/0301026, to appear in the Proceedings of TH2002 in Paris, July 2002;\\
K.~ Fujikawa and Masato Ishibashi, Nucl.Phys. B622 (2002) 115-140, {\it ibid} B605 (2001) 365-394, {\it ibid} B587 (2000) 419-441;\\
Kazuo Fujikawa, Masato Ishibashi, Hiroshi Suzuki, JHEP 0204 (2002) 046, Phys.Lett. B538 (2002) 197-201, hep-lat/0209007, Lattice2002(chiral), Phys.Rev. D67 (2003) 034506;\\
P. Hasenfratz, S. Hauswirth, K. Holland, T. Jorg, F. Niedermayer, Nucl.Phys.Proc.Suppl. 106 (2002) 751-756,
Nucl.Phys. B643 (2002) 280-320



\bibitem{xue99}
S.-S.~Xue, Phys.~Rev. D61 (2000) 054502.

\bibitem{xue00}
S.-S.~Xue, Nucl.~Phys.~B580 (2000) 365.

\bibitem{xuesm}
S.-S.~Xue, 
Nucl.~Phys.~B (Proc.~Suppl.) 94 (2001) 781.


\bibitem{11112}
S.-S.~Xue, Phys.~Rev. D64 (2001) 094504.

\bibitem{wilson}
K.~Wilson, in {\it New phenomena in subnuclear physics\/} 
(Erice, 1975) 
ed.\ A.~Zichichi (Plenum, New York, 1977).

\bibitem{xue90}
S.-S.~Xue, Phys.~Lett.~ B224,(1989),309, {\it ibid} B245,(1990),565, 
Z. Phys. C50,(1991),145,  
Il Nuovo Cimento A105,(1992),1749, {\it ibid} 105A (1992),1225.
 
\bibitem{wein}
S.~Weinberg, Phys.~Rev. 137 B672 (1965), {\it ibid} 130 776 (1963).

\bibitem{com}
D.~Lurie and A.J.~Macfarlane, Phys.~Rev. 136
(1964) B816.

\bibitem{smit82} L.~H.~Karsten and J.~Smit, Nucl.~ Phys. B144
(1978) 536.

\bibitem{preskill91}
J.~Preskill,  Ann.~Phys. NY, 210 (1991) 323.

\bibitem{witten} E.~Witten,  Phys.~Lett. B117
(1982) 324.

\bibitem{njl}
Y.~Nambu and G.~Jona-Lasinio, Phys. Rev.  122 (1961) 345.

\bibitem{lrmodel}
J.C.~Pati and A.~Salam, Phys.~Rev. D10 (1975) 275;\\
R.N.~Mohapatra and P.C.~Pati, {\it ibid} D11 (1975) 566;
D11 (1975) 2558;\\
G.~Senjanovi\'{c} and R.N.~Mohopatra, Phys.~Rev. D12 (1975) 1502;\\
R.N.~Mohopatra and G.~Senjanovi\'{c}, Phys.~Rev. D23 (1981) 165 and
Phys.~Rev.~Lett. 44 (1980) 912.

\bibitem{neu00}
S.-S.~Xue, Phys.~Lett.~B 377 (1996) 124; {\it ibid} B 398 (1997) 177;
Mod.~Phys.~Lett.~A Vol.~14 (1999) 2701; {\it ibid} Vol.~15 (2000) 1089;
the proceeding of Shizuoka's workshop on masses and mixings, Japan, March, 1997, p362.

\end{references}
\end{document}